\documentclass[aps,prd,10pt,
               reprint,
               notitlepage,
               preprintnumbers,numbers,sort&compress,nofootinbib,
               showpacs, 
               colorlinks,
               linkcolor=blue,
               citecolor=blue,
               amsmath, amssymb, 
               ]{revtex4-1}
\usepackage{graphicx,bm}
\usepackage{psfrag}
\usepackage{feynmp}
\usepackage{hyperref}   

\usepackage{microtype}

\newcommand{\fontchoice}[4]{#2}
\fontchoice{
  \usepackage[T1]{fontenc}
  \usepackage{ae}
  \usepackage{aecompl}
  \usepackage{lmodern}
  \usepackage{textcomp}
}{
  \usepackage[T1]{fontenc}
  \usepackage{ae}
  \usepackage{aecompl}
  \usepackage[sc]{mathpazo}
  \linespread{1.05}          
}{
  \usepackage[T1]{fontenc}
  \usepackage{ae}
  \usepackage{aecompl}
  
  \usepackage{euler}
  \linespread{1.05}          

}{
  \usepackage{minion}
  \usepackage{euler}
}

\usepackage{booktabs} 
\usepackage{enumitem} 
\setitemize{leftmargin=0pt,parsep=0pt,itemsep=0pt,
  listparindent=\parindent}
\setenumerate{leftmargin=0pt,labelsep=0.4ex,parsep=0pt,itemsep=0pt,
  listparindent=\parindent}

\renewcommand{\d}{\mathrm{d}}
\newcommand{\diff}[3][{}]{\frac{\mathrm{d}^{#1}{#2}}{\mathrm{d}{#3}^{#1}}}
\newcommand{\pdiff}[3][{}]{\frac{\partial^{#1}{#2}}{\partial{#3}^{#1}}}
\newcommand{\braket}[1]{\mathinner{\langle{#1}\rangle}}{\catcode`\|=\active
  \gdef\Braket#1{\left<\mathcode`\|"8000\let|\bravert {#1}\right>}}

\newcommand{\vect}[1]{\vec{\boldsymbol{#1}}}
\newcommand{\norm}[1]{\lVert{#1}\rVert}
\newcommand{\mat}[1]{\boldsymbol{#1}}
\newcommand{\e}{\mathrm{e}}     
\DeclareMathOperator{\Li}{Li}

\newcommand{\DFT}{\textsc{dft}}
\newcommand{\Chandra}{\textit{Chandra}}
\newcommand{\SPI}{\textsc{spi}}
\newcommand{\CGRO}{\textsc{crgo}}
\newcommand{\INTEGRAL}{\textsc{integral}}
\newcommand{\COMPTEL}{\textsc{comptel}}
\newcommand{\EGRET}{\textsc{egret}}
\newcommand{\WMAP}{\textsc{wmap}}
\newcommand{\QCD}{\textsc{qcd}}
\newcommand{\QED}{\textsc{qed}}
\newcommand{\ISM}{\textsc{ism}}
\newcommand{\DM}{\textsc{dm}}

\newcommand{\VHIM}{\textsc{vhim}}

\begin{document}
\preprint{LA-UR 09-06331}
\preprint{INT-PUB-10-054}
\begin{fmffile}{fmfprofile}

\title{The Electrosphere of Macroscopic ``Quark Nuclei'':\\
  A Source for Diffuse MeV Emissions from Dark
  Matter.}

\author{Michael McNeil Forbes}
\email[E-mail:~]{mforbes@alum.mit.edu}
\affiliation{Institute for Nuclear Theory, University of Washington, Box 351550,
  Seattle, Washington, 98195-1550, USA}

\author{Kyle Lawson}
\author{Ariel R. Zhitnitsky}
\affiliation{Department of Physics and Astronomy, University of
  British Columbia, Vancouver, BC, V6T 1Z1, Canada}
\date{\today}
\pacs{
  95.35.+d, 
  52.27.Aj, 
  78.70.Bj, 
  98.70.Rz, 
  }
\keywords{baryogenesis, antimatter, dark matter, dark antimatter,
  CDM, CCDM, x-rays, 511 keV, SPI, EGRET, CCO, strangelets, strange
  quark matter, density profiles, Thomas Fermi, Debye screening,
  strange stars, diffuse radiation, ISM, radiation mechanisms: non-thermal}

\begin{abstract}\noindent
  Using a Thomas-Fermi model, we calculate the structure of the
  electrosphere of the quark antimatter nuggets postulated to comprise
  much of the dark matter.  This provides a single self-consistent
  density profile from ultrarelativistic densities to the
  nonrelativistic Boltzmann regime that use to present
  microscopically justified calculations of several properties of the
  nuggets, including their net charge, and the ratio of MeV to 511~keV
  emissions from electron annihilation.  We find that the calculated
  parameters agree with previous phenomenological estimates based on
  the observational supposition that the nuggets are a source of
  several unexplained diffuse emissions from the Galaxy.  As no
  phenomenological parameters are required to describe these
  observations, the calculation provides another nontrivial
  verification of the dark-matter proposal. The structure of the
  electrosphere is quite general and will also be valid at the surface
  of strange-quark stars, should they exist.
\end{abstract}
\maketitle

\tableofcontents
\vspace{1em}                    
\section{Introduction}
\label{sec:introduction}\noindent
In this paper we explore some details of a testable and
well-constrained model for dark matter~\cite{Zhitnitsky:2002qa,
  Oaknin:2003uv, Zhitnitsky:2006vt, Forbes:2006ba, Forbes:2008uf} in
the form of quark matter as antimatter nuggets.  In particular, we
focus on physics of the ``electrosphere'' surrounding these nuggets:
It is from here that observable emissions emanate, allowing for the
direct detection of these dark-antimatter nuggets.

We first provide a brief review of our proposal in Sec.~\ref{sec:proposal}, then
describe the structure of the electrosphere of the nuggets using a Thomas-Fermi
model in Sec.~\ref{sec:electr-struct}.  This allows us to calculate the charge
of the nuggets, and to discuss how they maintain charge equilibrium with the
environment.  We then apply these results to the calculation of emissions from
electron annihilation in Sec.~\ref{sec:diffuse-1-20}, computing some of the
phenomenological parameters introduced in~\cite{Lawson:2007kp} required to
explain current observations.  The values computed in the present paper are
consistent with these phenomenologically motivated values, providing further
validation of our model for dark matter.  The present results concerning the
density profile of the electrosphere may also play an important role in the
study of the surface of quark stars, should they exist.

\section{Dark Matter as Dense Quark Nuggets}
\label{sec:proposal}\noindent
Two of the outstanding cosmological mysteries -- the natures of dark matter and
baryogenesis -- might be explained by the idea that dark matter consists of
compact composite objects (\textsc{cco}s)~\cite{Zhitnitsky:2002qa,
  Oaknin:2003uv, Zhitnitsky:2006vt, Forbes:2006ba, Forbes:2008uf} similar to
Witten's strangelets~\cite{Witten:1984rs}.  The basic idea is that these
\textsc{cco}s -- nuggets of dense matter and antimatter -- form at the same
\textsc{qcd} phase transition as conventional baryons (neutrons and protons),
providing a natural explanation for the similar scales $\Omega_{\DM} \approx
5\Omega_{B}$.  Baryogenesis proceeds through a charge separation
mechanism: both matter and antimatter nuggets form, but the natural \textsc{cp}
violation of the so-called $\theta$ term in \textsc{qcd}\footnote{If $\theta$ is
  nonzero, one must confront the so-called strong \textsc{cp} problem whereby
  some mechanism must be found to make the effective $\theta$ parameter
  extremely small today in accordance with measurements.  This problem remains
  one of the most outstanding puzzles of the Standard Model, and one of the most
  natural resolutions is to introduce an axion field.  (See the original papers
  \cite{Peccei:1977ur,*Weinberg:1978ma,*Wilczek:1978pj,
    Kim:1979if,*Shifman:1980if, Dine:1981rt,*Zhitnitsky:1980tq}, and recent
  reviews \cite{Srednicki:2002ww,*vanBibber:2006rb,*Asztalos:2006kz}.)  Axion
  domain walls associated with this field (or ultimately, whatever mechanism
  resolves the strong \textsc{cp} problem) play an important role in forming
  these nuggets, and may play in important role in their ultimate stability.
  See~\cite{Zhitnitsky:2002qa,Oaknin:2003uv,Dolgov:2009ix} for details.} --
which was of order unity $\theta\sim 1$ during the \QCD\ phase transition --
drives the formation of more antimatter nuggets than matter nuggets, resulting
in the leftover baryonic matter that forms visible matter today (see
\cite{Oaknin:2003uv} for details).  Note, it is crucial for our mechanism that
\textsc{cp} violation be able to drive charge separation: though not yet proven,
this idea may already have found experimental support through the Relativistic
Heavy Ion Collider (\textsc{rhic}) at Brookhaven \cite{Kharzeev:2007tn}, where
charge separation effects seem to have been observed~\cite{Voloshin:2010ut,
  Abelev:2010tx}

The mechanism requires no fundamental baryon asymmetry to explain the
observed matter/antimatter asymmetry.  Together with the observed
relation $\Omega_{\DM} \approx 5\Omega_{B}$
(see~\cite{Amsler:2008zz} for a review) we have
\begin{subequations}
  \label{eq:321}
 \begin{align}
    B_{\text{universe}} = 0 &= B_{\text{nugget}}
    + B_{\text{visible}}-\bar{B}_{\text{antinugget}}\\
    B_{\text{dark-matter}} &= B_{\text{nugget}} +
    \bar{B}_{\text{antinugget}} \approx 5B_{\text{visible}}
  \end{align}
\end{subequations}
where $B_{\text{universe}}$ is the overall asymmetry -- the total number of
baryons\footnote{Note that we use the term ``baryon'' to refer in general to
  anything carrying $U_{B}(1)$ baryonic charge.  This includes conventional
  colour singlet hadrons such as protons and neutrons, but also includes their
  constituents -- i.e. the quarks -- in other phases such as strange quark
  matter.}  \emph{minus} the number of antibaryons in the Universe -- and
$B_{\text{dark-matter}}$ is the total number of baryons \emph{plus} antibaryons
hidden in the dark-matter nuggets.  The dark matter comprises a baryon charge of
$B_{\text{nugget}}$ contained in matter nuggets, and an antibaryonic charge of
$\bar{B}_{\text{antinugget}}$ contained in antimatter nuggets.  The remaining
unconfined charge of $B_{\text{visible}}$ is the residual ``visible'' baryon
excess that forms the regular matter in our Universe today.  Solving
Eq.~(\ref{eq:321}) gives the approximate ratios
$\bar{B}_{\text{antinugget}}$:$B_{\text{nugget}}$:$B_{\text{visible}}\simeq $
3:2:1.

Unlike conventional dark-matter candidates, dark-matter/antimatter nuggets will
be strongly interacting, but macroscopically large, objects.  They do not
contradict any of the many known observational constraints on dark matter or
antimatter~\cite{Zhitnitsky:2006vt} for three reasons:
\begin{enumerate}
\item They carry a huge (anti)baryon charge $|B| \approx 10^{20}$ -- $10^{30}$,
  so they have an extremely tiny number density.\footnote{If the average nugget
    size ends up in the lower range, then the Pierre Auger observatory may
    provide an ideal venue for searching for these dark-matter candidates.}
  This explains why they have not been directly observed on earth.  The local
  number density of dark-matter particles with these masses is small enough that
  interactions with detectors are exceedingly rare and fall within all known
  detector and seismic constraints~\cite{Zhitnitsky:2006vt}.  (See also
  \cite{Herrin:2005kb, Abers:2007ji} and references therein.)
\item The nugget cores are a few times nuclear density $\rho \sim
  10$~GeV$/$fm$^3$, and thus have a size $R \sim 10^{-7}$--$10^{-3}$~cm.
  Their interaction cross section is thus small $\sigma/M \approx 4\pi R^2/M =
  10^{-13}$--$10^{-9}$~cm$^2$/g: well below the typical astrophysical and
  cosmological limits, which are on the order of $\sigma/M<1$~cm$^2$/g.
  Dark-matter--dark-matter interactions between these nuggets are thus
  negligible.
\item They have a large binding energy such that the baryonic matter
  in the nuggets is not available to participate in big bang
  nucleosynthesis (\textsc{bbn}) at $T \approx 1$~MeV.  In particular,
  we suspect that the core of the nuggets forms a superfluid with a gap
  of the order $\Delta \approx 100$~MeV, and critical temperature
  $T_{c} \sim \Delta/\sqrt{2} \approx 60$~MeV, as this scale provides
  a natural explanation for the observed photon to baryon ratio
  $n_{B}/n_{\gamma} \sim 10^{-10}$~\cite{Oaknin:2003uv},
  which requires a formation temperature of $T_{\text{form}} =
  41$~MeV~\cite{kolb94:_early_univer}.\footnote{At temperatures below
    the gap, incident baryons with energies below the gap would
    Andreev reflect rather than become incorporated into the nugget.}
\end{enumerate}

Thus, on large scales, the nuggets are sufficiently dilute that they behave as
standard collisionless cold dark matter (\textsc{ccdm}).  When the number
densities of both dark and visible matter become sufficiently high, however,
dark-antimatter--visible-matter collisions may release significant radiation and
energy.  In particular, antimatter nuggets provide a site at which interstellar
baryonic matter -- mostly hydrogen -- can annihilate, producing emissions that
should be observable from the core of our Galaxy of calculable spectra and
energy.  These emissions are not only consistent with current observations,
but naturally explain several mysterious diffuse emissions observed from the
core of our Galaxy, with frequencies ranging over 12 orders of magnitude.

Although somewhat unconventional, this idea naturally explains several
coincidences, is consistent with all known cosmological constraints,
and makes testable predictions.  Furthermore, this idea is almost
entirely rooted in conventional and well-established physics.  In
particular, there are no ``free parameters'' that can be -- or need to
be -- ``tuned'' to explain observations: In principle, everything is
calculable from well-established properties of \QCD\ and \QED.  In
practice, fully calculating the properties of these nuggets requires
solving the fermion many-body problem at strong coupling, so we have
generally resorted to ``fitting'' a handful of phenomenological
parameters from observations.  In this paper we examine the \QED\
physics of the electrosphere, providing a microscopic basis for some
of these parameters.  Once these parameters are determined, the model
makes unambiguous predictions about other processes ranging over more
than 10 orders of magnitude in scale.

The basic picture involves the antimatter nuggets -- compact cores of
nuclear or strange-quark matter (see Sec.~\ref{sec:core-structure})
surrounded by a positron cloud with a profile as calculated in
Sec.~\ref{sec:electr-struct}.  Incident matter will annihilate on
these nuggets producing radiation at a rate proportional to the
annihilation rate, thus scaling as the product
$\rho_{\textsc{v}}(\vect{r})\rho_{\textsc{dm}}(\vect{r})$ of the local visible and
dark-matter densities.  This will be greatest in the core of the
Galaxy.  To date, we have considered five independent observations of
diffuse radiation from the core of our Galaxy:
\begin{enumerate}
\item
  \href{http://smsc.cnes.fr/INTEGRAL/GP_instrument.htm}{S\textsc{pi}/\INTEGRAL}
  (the \SPI\ instrument on the \INTEGRAL\ satellite) observes 511~keV photons
  from positronium decay that is difficult to explain with conventional
  astrophysical positron sources~\cite{Knodlseder:2003sv, Beacom:2005qv,
    Yuksel:2006fj}.  Dark-antimatter nuggets would provide an unlimited source
  of positrons as suggested in~\cite{Oaknin:2004mn, Zhitnitsky:2006tu}.
\item
  \href{http://heasarc.gsfc.nasa.gov/docs/cgro/comptel/}{C\textsc{omptel}/\CGRO}
  (the \COMPTEL\ instrument on the Compton Gamma Ray Observatory (\CGRO)
  satellite) detects a puzzling excess of 1--20~MeV $\gamma$-ray radiation.  It
  was shown in \cite{Lawson:2007kp} that the direct $e^+e^-$ annihilation
  spectrum could nicely explain this deficit, but the annihilation rates were
  crudely estimated in terms of some phenomenological parameters.  In this paper
  we provide a microscopic calculation of these parameters
  (Eqs. (\ref{eq:branching}) and (\ref{eq:511norm})), thereby validating this
  prediction.
\item \href{http://chandra.harvard.edu/}{\Chandra} (the \Chandra\ x-ray
  observatory) observes a diffuse keV \textsc{x}-ray emission that greatly
  exceeds the energy from identified sources~\cite{Muno:2004bs}.
  Visible-matter/dark-antimatter annihilation would provide this energy.  It was
  shown in \cite{Forbes:2006ba} that the intensity of this emission is
  consistent with the 511~keV emission if the rate of proton annihilation is
  slightly suppressed relative to the rate of electron annihilation.  In
  Sec.~\ref{sec:nugg-charge-equil} we describe the microscopic nature of this
  suppression.
\item \href{http://heasarc.gsfc.nasa.gov/docs/cgro/egret/}{E\textsc{gret}/\CGRO}
  (the Energetic Gamma Ray Experiment Telescope aboard the \CGRO\ satellite)
  detects MeV to GeV gamma rays, constraining antimatter annihilation rates. It
  was shown in~\cite{Forbes:2006ba} that these constraints are consistent with
  the rates inferred from the other emissions.
\item \href{http://map.gsfc.nasa.gov/}{W\textsc{map}} (the Wilkinson Microwave
  Anisotropy Probe) has detected an excess of GHz microwave radiation -- dubbed
  the ``\WMAP\ haze'' -- from the inner $20^\circ$ core of our
  Galaxy~\cite{Finkbeiner:2003im, Finkbeiner:2004je,
    Hooper:2007gi,Dobler:2007wv}.  Annihilation energy not immediately released
  by the above mechanisms will thermalize, and subsequently be released as
  thermal bremsstrahlung emission at the eV scale.  In~\cite{Forbes:2008uf} it
  was shown that the predicted emission from the antimatter nuggets is
  consistent with, and could completely explain, the observed \WMAP\ haze.
\end{enumerate}
These emissions arise from the following mechanism: Neutral hydrogen
from the interstellar medium (\ISM) will easily penetrate into the
electrosphere, providing a source of electrons and protons.

The first and simplest process is the annihilation of the electrons through
positronium formation, producing 511~keV photons as discussed
in~\cite{Oaknin:2004mn, Zhitnitsky:2006tu}.  Note that this mechanism predicts
that, within the environment of the electrosphere, virtually all of the
low-energy emission should be characterized by a positronium decay spectrum,
including 25\% as a sharp 511~keV line from the decay of the singlet state and
the remaining 75\% as the broad three photon continuum resulting from the
triplet state.  As emphasized in~\cite{Lingenfelter:2009fk}, simply postulating
a dark-matter source of positrons does not suffice to explain the observed
spectrum characterized by $94\pm 4\%$ positronium annihilation: The positrons
must annihilate in the appropriate cool environment as provided by the
electrosphere of the nuggets.

Our proposal can thus easily explain the observed 511~keV radiation.  If we
assume that this process is the dominant source, then we can use this to
normalize the intensities of the other emissions.  A remarkable feature of this
proposal is that it then predicts the correct intensity for all of the other
observations, even though they span many orders of magnitude in frequency.

The second process is direct annihilation of the electrons on the positrons in
the electrosphere.  As discussed in appendix~\ref{sec:debye-1}, the electrons
are strongly screened by the positron background, and some fraction can
penetrate deep within the electrosphere.  There they can directly annihilate
with high-momentum positrons in the Fermi sea producing radiation up to 10~MeV
or so.  This process was originally discussed in~\cite{Lawson:2007kp} where the
ratio of direct MeV annihilation to 511~keV annihilation was characterized by
several phenomenological parameters chosen to fit the observations.  In
Sec.~\ref{sec:diffuse-1-20} we put this prediction on solid ground and show that
these parameter fits agree with the microscopic calculation based on the
electrosphere structure, which depends only on \QED.

The other radiation originates from the energy deposited by
annihilation of the incident protons in the core of the nuggets.
In~\cite{Forbes:2006ba} we argue that the protons will annihilate just
inside the surface of the core, releasing some 2~GeV of energy.
Occasionally this process will release GeV photons -- the rate of
which is consistent with the \EGRET/\CGRO\ constraints -- but most of
the energy will be transferred to strongly interacting components, and
ultimately about half will scatter down into $\sim 5$~MeV positrons
that stream out of the core.  (The $5$~MeV scale comes from the
effective mass of the photon in the medium which mediates the energy
exchange).  These positrons are accelerated by the strong electric
fields, and emit field-induced bremsstrahlung emission in the 10~keV
band -- a scale set by balancing the rate of emission with the local
plasma frequency (photons can only be emitted once the plasma
frequency is low enough).  The spectrum is also calculable: It is very
flat and similar to a thermal spectrum without the sharp falloff above
the temperature scale.  This is consistent with the \Chandra\
observations which cannot resolve the thermal falloff, but future
analysis might be able to distinguish between the two.

We shall shed some light on the interaction between the protons and
the core in Sec.~\ref{sec:nugg-charge-equil}, but cannot yet
perform the required many-body analysis to place all of this on a
strong footing as this would require a practical solution to
outstanding problems of high-density \QCD.

The remaining energy will thermalize within the nuggets, until an
equilibrium temperature of about $T\sim 1$~eV is
reached~\cite{Forbes:2008uf}.  Direct thermal emission at this scale
would be virtually impossible to see against the backgrounds, but the
spectrum -- calculable entirely from \QED\ -- extends to very low
frequencies, and the intensity of the emission in the microwave band
is just enough to explain the \WMAP\ haze~\cite{Forbes:2008uf}.

\section{Core Structure}
\label{sec:core-structure}\noindent
A full accounting of this nugget proposal requires a proper
description of the high-density phase found in the core of the
nuggets.  Unfortunately, a quantitative understanding of this phase
requires a practical solution to the notoriously difficult problem of
high-density \QCD.  The density of the core will be within several
orders of nuclear matter density $\sim 1-100$~GeV/fm$^{3}$.  This is
not high enough for the asymptotic freedom of \QCD\ may be used to
solve the problem perturbatively, so one must resort to
nonperturbative techniques such as the lattice formulation of \QCD.
Unfortunately, at finite density, the presence of the infamous sign
problem renders this approach exponentially expensive and it remains a
famously intractable problem.  As such, we cannot exactly quantify the
nature of the core and must constrain its properties from other
observations; we list the important properties in this section.

Fortunately, the observable emissions discussed in this paper result
primarily from the calculable physical processes in the electrosphere
of the nuggets, and are thus largely insensitive to the exact nature
of the core.  Nevertheless, the core structure must be addressed, and
it is possible that future developments concerning the properties of
high-density \QCD\ could rule out the feasibility of our nugget
proposal.

The first problem concerns the stability of the core.  All evidence suggests
that, in the absence of an external potential (such as the gravitational well of
a neutron star), nuclear matter will fragment into small nuclei with a baryon
number no larger than a few hundred.  This suggests one of two possibilities:
\begin{itemize}
\item The first possibility suggested by Witten~\cite{Witten:1984rs}
  is that a phase of strange-quark matter~\cite{Farhi:1984} becomes
  stable at high density.  In this case, the nuggets are simply
  strangelets and antistrangelets and the novel feature here is that
  the domain walls associated with strong $CP$ violation provide the
  required mechanism to condense enough matter to catalyze the
  formation of the strangelets before they evaporate.  (For a brief
  review, see~\cite{Madsen:2006} and references therein.)  

  Although strangelets have not yet been observed, the possibility of
  stable strange-quark mater has not yet been ruled out (see for
  example~\cite{Madsen:2004}).  This must be carefully reconciled with
  future astrophysical observations and constraints as it is
  conceivable that this possibility might be ruled out in the future.
  If the nuggets are a form of strange-quark matter, one must also
  consider the possibility of mixed phases as suggested
  in~\cite{Christiansen:1997} (see for example~\cite{Jaikumar:2006,
    Alford:2006} and references therein).  This would most likely have
  to be ruled out energetically to prevent the nuggets from
  fragmenting.
\item The second possibility is that the domain walls responsible for forming
  the nuggets at the \QCD\ phase transition become an integral part, providing a
  surface tension that holds the nuggets together, even in the absence of
  absolutely stable strange-quark matter.  The stability of this possibility has
  been discussed in detail in~\cite{Zhitnitsky:2002qa} and we shall not repeat
  these arguments here.  In this case, the core may be something more akin to
  dense nuclear matter as might be found in the core of a neutron star.
\end{itemize}

As discussed above, in order to explain the observations, the
following core properties are crucial to our proposal.  This provides
some insight into the required nature of the core:
\begin{enumerate}
\item The nuggets must be stable. Stable strange-quark matter would offer a nice
  explanation.  Otherwise, the structure of the core -- especially the surface
  -- must be considered in more detail.  This is a complicated problem that we
  do not presently know how to solve.
\item As discussed above, the formation of the objects must stop at
  $T=41$~MeV to explain the observed photon to baryon ratio.  This
  could be naturally explained by the order $100$~MeV pairing gaps
  expected in colour-superconducting strange-quark matter.  If the
  proposal turns out to be correct, then, this would provide a precise
  measurement of the pairing properties of high-density \QCD.  (The
  exact relationship between the pairing gap and the formation
  temperature will be quite nontrivial and require a detailed model
  of the formation dynamics.)  This favours a model of the core with
  strong pairing correlations.
\item In order to explain the Chandra data, our picture of the
  emission mechanism requires that the proton annihilations occur
  somewhat within the core, not immediately on the surface where there
  would be copious pion production.  The simplest explanation for this
  would be the presence of strong correlations in the core, delaying
  the annihilation until the protons penetrate a few hundred fm or so.
  Again, if the core is a colour superconductor, then the pairing
  correlations could explain this, but a detailed calculation is
  needed to make sure.

  An argument supporting the presence of such correlations in a dense
  environment is the observation that the annihilation cross section
  of an antiproton on a heavy nuclei is many times smaller than the
  vacuum $p\bar{p}$ and $n\bar{n}$ annihilation rates suggest.
  Indeed, it is expected that antiproton/nuclei bound states may
  persist with a lifetime much longer than the vacuum annihilation
  rates predicts (see~\cite{Mishustin:2004xa, Larionov:2008wy}).
  This annihilation suppression should become even stronger with
  increasing density.
\end{enumerate}
Thus, the hypothesis of strange-quark matter simplifies the picture
quite a bit, but we do not yet see a way to directly link this
hypothesis with the existence or stability of the nuggets.

\section{Electrosphere Structure}
\label{sec:electr-struct}\noindent
Here we discuss the density profile of the electron cloud surrounding
extremely heavy macroscopic ``nuclei''\kern-1ex.  We present a type of
Thomas-Fermi analysis including the full relativistic electron
equation of state required to model the relativistic regime close to
the nugget core.  We consider here the limit when the temperature is
much smaller than the mass $T \ll 511$~MeV.  The solution for
higher temperatures follows from similar techniques with fewer
complications.

The observable properties of the antimatter nuggets discussed so
far~\cite{Zhitnitsky:2002qa, Oaknin:2003uv, Zhitnitsky:2006vt,
  Forbes:2006ba, Lawson:2007kp, Forbes:2008uf} depend on the existence
of a nonrelativistic ``Boltzmann'' regime with a density dependence
$n(r) \sim (r - r_{\textrm{B}})^{-2}$ (see appendix 1 of
Ref.~\cite{Forbes:2008uf} for details).  This region plays an
important role in explaining the \WMAP\ haze \cite{Forbes:2008uf} as
well as in the analysis of the diffuse 511~keV emissions
\cite{Oaknin:2003uv}.  The techniques previously used, however, were
not sufficient to connect this nonrelativistic ``Boltzmann'' regime
to the relativistic regime through a self-consistent solution
determined by parameters $T, \mu, m_{\text{e}}$.

The main point of this section is to put the existence of a sizable Boltzmann
regime on a strong footing, and to calculate some of the previously estimated
phenomenological parameters that depend sensitively on density profile.  These
may now be explicitly computed from \textsc{qed} using a justified Thomas-Fermi
approximation to reliably account for the many-body physics.  We show that a
sizable Boltzmann regime exists for all but the smallest nuggets, which are
ruled out by lack of terrestrial detection observation.  We also address the
question of how the nuggets achieve charge equilibrium
(Sec.~\ref{sec:nugg-charge-equil}), discussing briefly the charge-exchange
mechanism, and determining the overall charge of the nuggets (see
Table~\ref{tab:Antinugget}).

\subsection{Thomas-Fermi Model}
\noindent
To model the density profile of the electrosphere, we use a Thomas-Fermi model
for a Coulomb gas of positrons.  This is derivable from a density functional
theory (see Appendix~\ref{sec:dens-funct-theory}) after neglecting the exchange
contribution, which is suppressed by the weak coupling $\alpha$.  The
electrostatic potential $\phi(\vect{r})$ must satisfy the Poisson equation
\begin{equation}
  \label{eq:Poisson}
  \nabla^2 \phi(\vect{r}) = -4\pi e n(\vect{r}).
\end{equation}
where $e n(\vect{r})$ is the charge density.  Outside of the nugget core, we
express everything in terms of the local effective chemical potential
\begin{equation}
  \label{eq:mu}
  \mu(\vect{r}) = - e\phi(\vect{r}),
\end{equation}
and express the local charge density through the function $q(\vect{r}) =
en[\mu(\vect{r})]$ where $n[\mu]$ contains all of the information about the
equation of state.

As with the Thomas-Fermi model of an atom, the self-consistent solution will be
determined by the charge density of the core (``nucleus'').  We shall simply
implement this as a boundary condition at the nugget core boundary at radius
$r=R$, and thus only consider the region $r>R$.  The resulting solution may thus
be expressed in terms of the equations
\begin{subequations}
  \label{eq:TF}
  \begin{gather}
    \label{eq:DEQr}
    \begin{aligned}
      \nabla^{2}\mu(\vect{r}) &= 4\pi \alpha n[\mu(\vect{r})], &
      \epsilon_{\vect{p}} &= \sqrt{\norm{\vect{p}}^2+m^2},
    \end{aligned}\\
    \label{eq:n_mu}
    \!n[\mu] = 
    2\!\int\!\!\frac{\d^{3}\vect{p}}{(2\pi)^3}\!\left[
      \frac{1}{1+\e^{(\epsilon_{\vect{p}}-\mu)/T}} -
      \frac{1}{1+\e^{(\epsilon_{\vect{p}}+\mu)/T}}\right]\!,
  \end{gather}
\end{subequations}
with the appropriate boundary conditions at $r=R$ and $r=\infty$.
In~(\ref{eq:n_mu}) we have explicitly included both particle and antiparticle
contributions, as well as the spin degeneracy factor, and have used modified
Planck units where $\hbar = c = 4\pi\epsilon_{0} = 1$ so that $e^2 = \alpha$ and
energy, momentum, inverse distance, and inverse time are expressed in eV.

We assume spherical symmetry so that we may write
\begin{equation}
  \label{eq:RadialLaplacian}
  \nabla^2\mu(r) = \frac{1}{r^2}\diff{}{r}r^2\diff{}{r}\mu(r)
  =\mu''(r) + 2\mu'(r)/r.
\end{equation}
Close to the surface of the nuggets, the radius is sufficiently large compared
with the relevant length scales that the curvature term $2\mu'(r)/r$ may be
neglected.  We shall call the resulting approximation
``one-dimensional''\kern-1ex.  The resulting profile does not depend on the
size of the nuggets and remains valid until the electrosphere extends to a
distance comparable to the radius of the nugget, at which point the full
three-dimensional form will cut off the density.  When applied to the context of
strange stars~\cite{Alcock:1986hz, Kettner:1994zs, Cheng:2003le, Usov:2004kj},
the one-dimensional approximation will be completely sufficient.  See
also~\cite{Heiselberg:1993} where a Thomas-Fermi calculation is used to
determine the charge distribution inside a strange-quark nugget.

To help reason about the three-dimensional equation, we note
that~(\ref{eq:DEQr}) may be expressed as
\begin{equation}
  \label{eq:DEQx}
  \mu''(x) = x^{-4} 4\pi \alpha n[\mu(x)]
\end{equation}
where $x=1/r$.  A nice property of this transformation is that
$\mu(x)$ must be a convex function if the charge density has the same
sign everywhere.

\subsubsection{Boundary Conditions}
\noindent
The physical boundary condition at the origin follows from smoothness of the
potential.  By combining this with a model of the charge distribution in the
core and an appropriate long-distance boundary-condition, one could in principle
model the entire distribution of electrons throughout the nugget.

In the nugget core, however, beta-equilibrium essentially
establishes a chemical potential on the order of $10$ --
$100$~MeV~\cite{Alcock:1986hz} that depends slightly on the exact
equation of state for the quark-matter phase (which is not
known). Thus, we may simply take the boundary condition as $\mu(R) =
\mu_{R} \approx 25$~MeV.  Our results are not very sensitive to the
exact value, though if less that 20~MeV, then this acts as a cutoff
for the direct $e^+e^-$ emission discussed in
Sec.~\ref{sec:diffuse-mev-scale}.

The formal difficulty in this problem is properly formulating the
long-distance boundary conditions.  At $T=0$, the large-distance
boundary condition for nonrelativistic systems is clear:
$n(r\rightarrow\infty) = 0$.  From (\ref{eq:DEQx}) we see that
$\mu(x)$ is linear with slope $\mu'(x)=-eQ$ where $Q$ is the
overall charge of the system (ions with a deficiency of
electrons/positrons are permitted in the theory; see for
example~\cite{LL3:1977}):
\begin{equation*}
  eQ(r) = \int_{0}^{r}\!\!\d{\tilde{r}}\; 4\pi\alpha n(\tilde{r}) \tilde{r}^2
  = \int_{0}^{r}\!\!\d{\tilde{r}}\;\left(\tilde{r}^2\mu'(\tilde{r})\right)'
  = r^2\mu'(r).
\end{equation*}
At finite temperature, however, this type of boundary-condition is not
appropriate.  Instead, one must consider how equilibrium is established.

In a true vacuum, a finite temperature nugget will ``radiate'' the loosely bound
outer electrosphere until the electrostatic potential is comparable to the
temperature $eQ/r_{*} \sim T$.  At this radius, radiation becomes exponentially
suppressed.  Suppose that the density at this radius is $n(r_{*})$.  We can
estimate an upper bound for the evaporation rate as $4\pi r_{*}^2 v n$ where $v
\sim \sqrt{T/m}$.\footnote{The Boltzmann averaged velocity is lower $\braket{v}
  = \sqrt{T/(2\pi m)}$, and cutting off the integral properly will lower this
  even more, so this gives a conservative upper bound on the evaporation rate.}

This needs to be compensated by rate of charge deposition from the
surrounding plasma which can be estimated as $4\pi R^2
v_{\textsc{n}^{+},\ISM} n_{\textsc{n}^{+},\ISM}$ where
$v_{\textsc{n}^{+},\ISM}\sim 10^{-3}c$ is the typical relative speed
of the nuggets and charged components in the Inter-Stellar Medium
(\ISM) and $4\pi R^2$ is approximately the cross section for
annihilation on the core (see Sec.~\ref{sec:nugg-charge-equil}).
The density $n_{\textsc{n}^{+},\ISM}$ of charged components in the
core of the Galaxy is typically $10^{-1}$ to $10^{-2}$ of the total
density $n_{\ISM} \sim 1$~cm$^{-3}$.  Thus, once the density falls
below
\begin{equation}\label{eq:n_rad}
  n < n_{\text{rad}} \sim 
  \frac{R^2}{r^2} \sqrt{\frac{m}{T}} v_{\textsc{n}^+,\ISM}
  n_{\ISM}10^{-2}
\end{equation}
charge equilibrium can easily be established.  This allows us to formulate the
long-distance boundary conditions by picking a charge $Q$ and outer radius
$r_{*}$ such that $eQ/r_{*} \sim T$, and $n(r_{*}) \sim n_{\text{rad}}$,
establishing both the typical charge of the nuggets as well as the outer
boundary condition for the differential
equation~(\ref{eq:TF}).\footnote{Strictly speaking, at finite $T$, the equations
  do not have a formal solution with a precise total charge because there is
  always some density for $\mu >0$. Practically, once $m - \mu\ll T$, the
  density becomes exponentially small and the charge is effectively
  fixed.}

Note that the Thomas Fermi approximation is not trust-worthy in these
regimes of extremely low density, but the boundary condition suffices
to provide an estimate of the charge of the nuggets: were they to be
less highly charged, then the density at $r_{*}$ would be sufficiently
high that evaporation would increase the charge; were they more highly
charged, then the evaporation rate would be exponentially suppressed,
allowing charge to accumulate.
\clearpage\noindent
Thus, for the antimatter nuggets, the density profile is determined by
system~(\ref{eq:TF}) and the boundary conditions
\begin{subequations}
  \label{eq:BC}
  \begin{gather}\label{eq:BCa}
    \mu(R) = \mu_{R} \sim 25\text{ MeV},\\
    \label{eq:BCb}
    \begin{aligned}
      eQ/r_{*} &= T \sim 1\text{ eV}, &
      n(r_{*}) &= n_{\text{rad}}(r_*).
    \end{aligned}
  \end{gather}
\end{subequations}

\subsection{Profiles}
\label{sec:profiles}\noindent
In principle, one should also allow the temperature to vary.  However, tThe rate
of radiation in the Boltzmann regime is suppressed by almost six orders of
magnitude with respect to the black-body rate; meanwhile, the density of the
plasma $n_{\textrm{B}}\sim (mT)^{3/2}$ is quite large~\cite{Forbes:2008uf}.
Thus, the abundance of excitations ensures that the thermal conductivity is high
enough to maintain an essentially constant temperature throughout the
electrosphere.

\begin{table}[b]
  \begin{tabular}{l@{\hspace{2em}}r@{\hspace{2em}}r@{\hspace{2em}}r}
    \toprule
     $\hphantom{1}B$ & $n_{\text{core}}\hphantom{{}^{-3}}$ & $R$\hphantom{~cm} &
     $Q\hphantom{10e}$\\
     \midrule
    $10^{20}$ & $100\text{ fm}^{-3}$ & $10^{-7}$ cm & $5\times 10^{1}e$\\
    & $1\text{ fm}^{-3}$ & $5\times 10^{-7}$ cm & $2\times 10^{2}e$\\
    $10^{24}$ & $100\text{ fm}^{-3}$ & $2\times 10^{-6}$ cm & $10^{3}e$\\
    & $1\text{ fm}^{-3}$ & $10^{-5}$ cm & $4\times 10^{3}e$\\
    $10^{33}$ & $100\text{ fm}^{-3}$ & $2\times 10^{-3}$ cm & $10^{6}e$\\
    & $1\text{ fm}^{-3}$ & $10^{-2}$ cm & $4\times 10^{6}e$\\
    \bottomrule
  \end{tabular}
  \caption{\label{tab:Antinugget}
    Antimatter nugget properties: the core radius $R$ and
    total electric charge $Q$ for nuggets with two different assumed core
    densities $n_{\text{core}}$ and three different antibaryon
    charges $B$.  These correspond to the profiles shown in
    Fig.~\ref{fig:antinugget_profiles}.}
\end{table}

\begin{figure}[b]
  \begin{center}
    \includegraphics[width=\columnwidth]{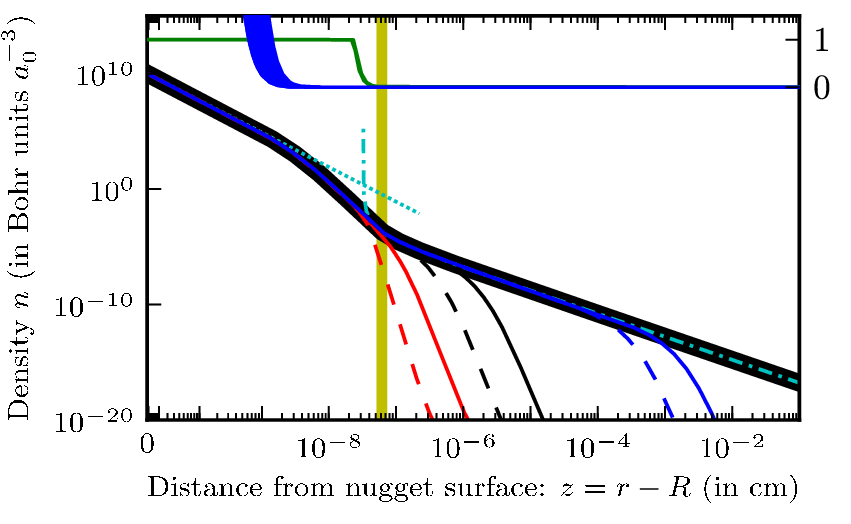}\\
    \includegraphics[width=\columnwidth]{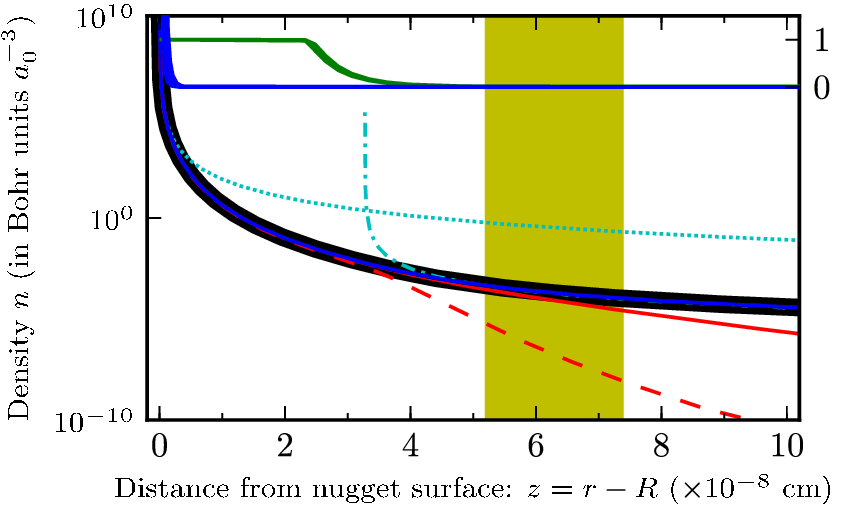}
    \caption{\label{fig:antinugget_profiles} Density
      profile of antimatter nuggets (bottom plot is a zoom).  The
      thick solid line is the one-dimensional approximation neglecting
      the curvature of the nugget.  Six profiles are shown descending
      from this in pairs of thin solid and dashed curves.  From left
      to right, each pair has fixed baryon charge $B=10^{20}$ (red),
      $10^{24}$ (black) and $10^{33}$ (blue) respectively. The solid
      curves represent nuclear density cores while the dashed curves
      represent 100 times nuclear density.  The light shaded (yellow)
      regions correspond to the nonrelativistic Boltzmann regime
      where the Boltzmann approximation~(\ref{eq:Boltzmann})
      (dash-dotted (cyan) line) is valid.  Only the $B=10^{20}$
      profiles visibly depart from the one-dimensional approximation
      in this regime.  The two upper curves in the top plot use the
      scale on the right and are the  annihilation rates $\Gamma_{Ps}$
      (right (green) curve) and $\Gamma_{\text{dir}}$ (left (blue) curve)
      normalized to the saturated value $\Gamma_{Ps}(\mu=\infty) = 4v
      q^3/(3\pi m^2\alpha^2)$~(\ref{eq:gamma_Ps}).  These curves
      comprise a range of cutoffs $q \sim m\alpha$ such the
      positronium annihilation rates~(\ref{eq:gamma_Ps}) vary by 10\%,
      and a range of incoming velocities $10^{-3}c < v < 10^{-2}c$.
      The scaling of the abscissa in the upper figure is $\ln(z +
      z_{\textsc{ur}})$ where $z_{\textsc{ur}} \approx 5\times
      10^{-11}$ so that the $T=0$ ultrarelativistic
      approximation~(\ref{eq:UR}) is linear (finely dotted (cyan)
      line).  The abscissa are linearly spaced in the bottom figure.}
  \end{center}
\end{figure}

One can now numerically solve the system~(\ref{eq:TF})
and~(\ref{eq:BC}).  We shall consider six cases: total (anti)baryon
charge $B\in\{10^{20}, 10^{24}, 10^{33}\}$ and quark-matter densities
$n_{\text{core}} \in \{1, 100\}\text{ fm}^{-3}$.  The resulting
profiles are plotted in
Fig.~\ref{fig:antinugget_profiles}.\footnote{
  \label{foot:cumberbatch}
  Note that the ultrarelativistic approximation employed in
  \cite{Cumberbatch:2006bj} discussing nonrelativistic physics not
  only overestimates the density by 3 orders of magnitude in the
  Boltzmann regime, but also has a different $z$-dependence
  [(\ref{eq:UR}) verses (\ref{eq:Boltzmann})]: The ultrarelativistic
  approximation is \emph{not} valid for $\mu < m$ where most of the
  relevant physical processes take place.}  As a reference, we also
plot the ``one-dimensional'' approximation obtained numerically by
neglecting the curvature term in~(\ref{eq:RadialLaplacian}) as well as
two analytic approximations: One for the ultrarelativistic
regime~\cite{Alcock:1986hz, Kettner:1994zs, Usov:2004iz}\footnote{For
  $T\neq 0$, see appendix~\ref{sec:ultrarelat-regime}.} where $n[\mu]
\approx \mu^3/3\pi^2$:
\begin{align}\label{eq:UR}
  n_{\textsc{ur}}(z) &\approx 
  \frac{\mu_{R}^3}{3\pi^2(1 + z/z_{\textsc{ur}})^3}, &
  z_{\textsc{ur}} &= \mu_{R}^{-1}\sqrt{\frac{3\pi}{2\alpha}},
\end{align}
and one for the nonrelativistic Boltzmann regime~\cite{Forbes:2008uf}
where $n[\mu] \propto \e^{\mu/T}$:
\begin{equation}\label{eq:Boltzmann}
  n_{\textrm{B}}(z) = \frac{T}{2\pi\alpha}\frac{1}{(z + z_{\textrm{B}})^2}.
\end{equation}
Since we have the full density profiles, we fit $z_{\textrm{B}}$ here so that
the this approximation matches the numerical solution at $n_{\textrm{B}} =
(mT/2\pi)^{3/2}$, (a slightly better approximation than used
in~\cite{Forbes:2008uf}).  

All of our previous estimates about physical properties of the
antimatter nuggets -- emission of radiation, temperature etc. -- have
been based on these approximations using the region indicated in
Fig.~\ref{fig:antinugget_profiles}, it is clear that they work very
well, \emph{as long as these regimes exist}.  The potential problem is
that the nuggets could be highly charged or sufficiently small that
the one-dimensional Boltzmann regime fails to exist with the density
rapidly falling through the relevant density scales.  We can see from
the results in Fig.~\ref{fig:antinugget_profiles} that the
nonrelativistic region exists for all but the smallest nuggets: As
long as $B>10^{20}$ -- as required by current detector
constraints~\cite{Forbes:2006ba} -- then the one-dimensional
approximation suffices to calculate the physical properties.

The annihilation rates discussed in Sec.~\ref{sec:diffuse-1-20}
have also been included in Fig.~\ref{fig:antinugget_profiles} to show
that these effects only start once the densities have reached the
atomic density scale, which is higher than the Boltzmann regime.  Thus
these results are also insensitive to the size of the nugget and the
domain-wall approximation suffices.  It is clear that one must have a
proper characterization of the entire density profile from
nonrelativistic through to ultrarelativistic regimes in order to
properly calculate the emissions: one cannot use simple analytic
forms.

We now discuss how charge equilibrium is established, and then
apply our results to fix the relative normalization between the
diffuse 1 -- 20~MeV emissions and the 511~keV emissions from the core
of our Galaxy.  As we shall show, the relative normalization is now
firmly rooted in conventional physics, and in agreement with the
previous phenomenological estimate~\cite{Lawson:2007kp}, providing
another validation of our theory for dark matter.

\subsection{Nugget Charge Equilibrium}
\label{sec:nugg-charge-equil}\noindent
In order to determine the effective charge of the nuggets we must
consider how equilibrium is obtained.  For the matter nuggets,
equilibrium is established through essentially static equilibrium with
the surrounding \ISM\ plasma, however, for the antimatter nuggets, no
such static equilibrium can be achieved.  Instead, one must consider
the dynamics of the following charge-exchange processes:
\begin{enumerate}
\item Deposition of charge via interaction and annihilation of neutral
  \ISM\ components (primarily neutral hydrogen).
\item Deposition of charge via interaction and annihilation of ionized
  \ISM\ components (electrons, protons, and ions).
\item Evaporation of positrons from the antinugget's surface.
\end{enumerate}

First we consider impinging neutral atoms or molecules with velocity
$v_{\textsc{n}, \ISM} \sim 10^{-3}c$.  Even if the antinugget is charged, these
are neutral and will still penetrate into the electrosphere.  Here the electrons
will annihilate or be ionized leaving a positively charge nucleus with energy
$T_{\ISM} \sim \tfrac{1}{2}m_{\textsc{n}} v_{\textsc{n},\ISM}^2$.  As the
electrosphere consists of positrons, this charge cannot be screened, so the
nucleus will accelerate to the core.  At the core the nucleus either will
penetrate and annihilate, resulting in the diffused x-ray emissions discussed
in~\cite{Forbes:2006ba} and providing the heat to fuel the microwave
emissions~\cite{Forbes:2008uf}, or will bounce off of the surface due to the
sharp quark-matter interface.\footnote{It is well known from quantum mechanics
  that any sufficiently sharp transition has a high probability of reflection of
  low energy particles.}

\begin{figure}[b]
  \begin{center}
    \setlength{\unitlength}{\columnwidth}
    \def\grey{(0.5,,0.5,,0.5)}
    \begin{fmfgraph*}(0.8,0.35)
      \fmfstraight
      \fmfleftn{i}{7}
      \fmfrightn{o}{7}
      \fmfv{label=$u$}{i7}
      \fmfv{label=$d$}{i6}
      \fmfv{label=$u$}{i5}
      \fmfv{label=$u$}{o7}
      \fmfv{label=$d$}{o6}
      \fmfv{label=$d$}{o5}
      \fmfv{label=$\bar{B}$}{i2}
      \fmfv{label=$\bar{B}$}{o2}
      \fmf{fermion,right=0.2}{i7,o7}
      \fmf{fermion,right=0.15}{i6,o6}
      \fmf{phantom,right=0.1}{i5,o5}
      \fmf{phantom}{o3,i3}
      \fmf{fermion}{o2,i2}
      \fmffreeze
      \fmfipath{p[]}                    
      \fmfiset{p10}{vpath (__i5,__o5)}
      \fmfiset{p11}{vpath (__o3,__i3)}
      \fmfiset{p1}{vpath (__i7,__o7)}
      \fmfiset{p2}{vpath (__i6,__o6)}
      \fmfiset{p3}{subpath (0,0.4)*length(p10) of p10 ..
        subpath (0.6,1)*length(p11) of p11}
      \fmfiset{p4}{subpath (0,0.4)*length(p11) of p11 ..
        subpath (0.6,1)*length(p10) of p10}
      \fmfiset{p5}{vpath (__o2,__i2)}
      \fmfi{fermion}{p3}
      \fmfi{fermion}{p4}
      \fmfi{fermion}{vpath (__o2,__i2) shifted ((0,0.1111h))}
      \fmfi{fermion}{vpath (__o2,__i2) shifted ((0,0.0555h))}
      \fmfi{fermion}{vpath (__o2,__i2) shifted ((0,-0.0555h))}
      \fmfi{fermion}{vpath (__o2,__i2) shifted ((0,-0.1111h))}
      \fmfset{curly_len}{1.5mm}
      \fmfcmd{
        pair a[], k[];
        a1 = (0.2,0.3); k1 = (1,2);
        a2 = (0.8,0.6); k2 = (1,2);
        a3 = (0.6,0.4); k3 = (3,4);
        a4 = (0.6,0.3); k4 = (4,3);
        a5 = (0.2,0.7); k5 = (4,4);
        a6 = (0.1,0.8); k6 = (3,5);
        a7 = (0.8,0.1); k7 = (2,5);
        for n = 1 upto 7:
          idraw ("gluon,fore=\grey,width=0.1thin,rubout=15",
          point (xpart a[n])*length(p[xpart k[n]]) of p[xpart k[n]] --
          point (ypart a[n])*length(p[ypart k[n]]) of p[ypart k[n]]);
        endfor;
      }
    \end{fmfgraph*}\\[5em]
    \begin{fmfgraph*}(0.8,0.35)
      \fmfstraight
      \fmfleftn{i}{7}
      \fmfrightn{o}{7}
      \fmfv{label=$u$}{i7}
      \fmfv{label=$d$}{i6}
      \fmfv{label=$u$}{i5}
      \fmfv{label=$u$}{o7}
      \fmfv{label=$d$}{o6}
      \fmfv{label=$u$}{o5}
      \fmfv{label=$\bar{B}$}{i2}
      \fmfv{label=$\bar{B}$}{o2}
      \fmf{fermion,right=0.2}{i7,o7}
      \fmf{fermion,right=0.15}{i6,o6}
      \fmf{fermion,right=0.1}{i5,o5}
      \fmf{fermion}{o3,i3}
      \fmf{fermion}{o2,i2}
      \fmffreeze
      \fmfipath{p[]}                    
      \fmfiset{p1}{vpath (__i7,__o7)}
      \fmfiset{p2}{vpath (__i6,__o6)}
      \fmfiset{p3}{vpath (__i5,__o5)}
      \fmfiset{p4}{vpath (__o3,__i3)}
      \fmfiset{p5}{vpath (__o2,__i2)}
      \fmfi{fermion}{vpath (__o2,__i2) shifted ((0,0.1111h))}
      \fmfi{fermion}{vpath (__o2,__i2) shifted ((0,0.0555h))}
      \fmfi{fermion}{vpath (__o2,__i2) shifted ((0,-0.0555h))}
      \fmfi{fermion}{vpath (__o2,__i2) shifted ((0,-0.1111h))}
      \fmffreeze
      \fmfset{curly_len}{1.5mm}
      \fmfcmd{
        pair a[], k[];
        a1 = (0.2,0.3); k1 = (1,2);
        a2 = (0.8,0.6); k2 = (1,2);
        a3 = (0.6,0.4); k3 = (3,4);
        a4 = (0.3,0.6); k4 = (3,4);
        a5 = (0.2,0.3); k5 = (4,5);
        a6 = (0.25,0.8); k6 = (3,5);
        a7 = (0.8,0.1); k7 = (2,5);
        for n = 1 upto 7:
          idraw ("gluon,fore=\grey,width=0.1thin,rubout=15",
          point (xpart a[n])*length(p[xpart k[n]]) of p[xpart k[n]] --
          point (ypart a[n])*length(p[ypart k[n]]) of p[ypart k[n]]);
        endfor;
      }
    \end{fmfgraph*}
    \caption{\label{fig:exch} Charge-exchange process (top diagram): A
      charged incoming proton (upper left) exchanges an up quark $u$
      for a down quark $d$ with the antimatter nugget $\bar{B}$,
      reflecting as a neutron (upper right).  This is Zweig suppressed
      relative to the simple reflection process illustrated below,
      however, the neutron can escape from the system whereas the
      reflected proton will be trapped by the electric fields.  This
      will enhance the overall rate of charge-exchange reactions
      because the proton will continue to react again and again until
      either the charge exchange occurs, or it eventually annihilates.
      In order to explain the relative intensities of the observed
      511~keV and diffuse x-ray emissions from the core of our
      Galaxy~\cite{Forbes:2006ba}, the ratio of the charge-exchange
      rate to the annihilation rate must be $f\sim 0.1$--$0.5$.  To
      calculate this from microscopic physics, however, requires
      details of the nugget's surface and equation of state
      that are not yet known.  Perhaps some model-insensitive
      estimates could be made, which would provide a highly
      nontrivial test of our theory based on microscopic physics.}
  \end{center}
\end{figure}
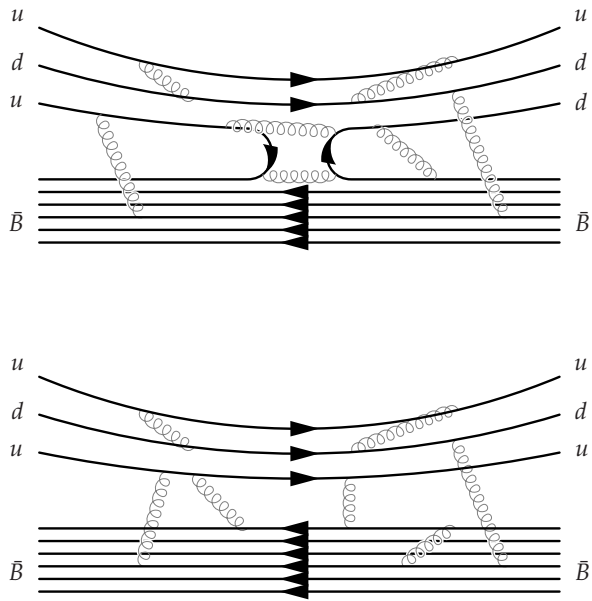

Note that, if the nuggets are sufficiently charged such that kinetic
energy of the incoming nucleus at the point of ionization is smaller
than the electrostatic energy,
\begin{equation}
  \label{eq:CoulombBarrier}
  T_{\ISM} \sim \tfrac{1}{2}m_{\textsc{n}} v_{\textsc{n},\ISM}^2 \sim
  \text{eV}
  \ll \frac{eQ}{r} \sim \text{keV} \sim 10^{7}\text{ K},
\end{equation}
then the charged nucleus will be unable to escape and will return to the core to
either annihilate or undergo a charge-exchange reaction to become neutral, after
which it may leave unimpeded by the electric field.  For nuggets with $B\sim
10^{20}$, $10^{24}$, and $10^{33}$ respectively, this critical charge is $Q/e
\gg 1$, $10^{2}$, and $10^{4}$ respectively.  As shown in
Table~\ref{tab:Antinugget}, the charge established through evaporation greatly
exceeds this critical charge: The charged nucleus will not be able to escape the
antinugget unless it undergoes a charge-exchange reaction (see
Fig.~\ref{fig:exch}).
 
For a single proton, such a charge-exchange reaction consists of an up
quark being replaced by a down quark at the surface of the nugget,
converting the proton to a neutron which can then escape.  This
process involves the exchange of a quark-antiquark pair and is a
strong interaction process, but suppressed by the Zweig (\textsc{ozi})
rule.  The overall ratio of charge-exchange to annihilation reactions
is amplified by a finite probability of reflection from the core
boundary: this results in multiple bounces before annihilation.

A better understanding of the details concerning the interaction
between the proton and the quark-matter boundary is required to
quantitatively predict the ratios of the rate of charge-exchange to
the rate of annihilation and thus to confirm or rule out the
suppression factor $f \sim 0.1$--$0.5$ required to explain the
relative strengths of the observed 511~keV and diffuse $\sim 10$~keV
x-ray emissions seen from the core of the Galaxy~\cite{Forbes:2006ba}.
This requires an estimate of the rates of reflection, annihilation,
and charge-exchange; the elasticity of collisions; and the energy loss
through scattering.

We emphasize that the corresponding calculations do not require any
new physics: everything is rooted firmly in \textsc{qed} and
\textsc{qcd}.  They do, however, require solving the many-body physics
of the strong interaction including the incident nucleons and the
quark antimatter interface (the phase of which may be quite
complicated).  A full calculation is thus very difficult, requiring
insight into high-density \textsc{qcd}: estimates can probably be made
using standard models of nuclear matter for the core, but are beyond
the scope of the present paper.

In any case, the nucleus will certainly deposit its charge on the
antinugget, and one may neglect the interactions with neutral \ISM\ 
components for the purpose of establishing charge neutrality.

Instead, we must consider the charged components.  Using the same
argument~(\ref{eq:CoulombBarrier}) one can see that the Coulomb
barrier of the charged nuggets will be high enough to prevent
electrons from reaching the electrosphere in all but the very hot
ionized medium (\VHIM), which occupies only a small fraction of the
\ISM\ in the core (see \cite{Ferriere:2007rz} for example).  Protons
however, will be able to reach the core to annihilate with a
cross section $\sim 4\pi R^2$ (this is not substantially affected by
the charge).  Thus, positive charge can be deposited at a rate $\sim
4\pi R^2 v_{\textsc{n}^+,\ISM}n_{\textsc{n}^+,\ISM}$ where
$n_{\textsc{n}^+,\ISM}$ is the density of the ionized components in
the \ISM, which is typically $10^{-1}$ to $10^{-2}$ of the total
density $n_{\ISM} \sim 1$~cm$^{-3}$

This rate of positive charge accumulation must match the evaporation
rate of the positrons from the electrosphere (see
Eq.~(\ref{eq:n_rad})), giving the boundary conditions~(\ref{eq:BCb}),
and the resulting charges summarized in Table~\ref{tab:Antinugget}.
(These are consistent with the estimates in~\cite{Abers:2007ji} and
the universal upper bound~\cite{Madsen:2008}.)

Note that the rate of evaporation is much less than the rate of proton
annihilation and carries only $\sim 1$~eV of every per particle
compared to the nearly 2~GeV of energy deposited.  Thus, evaporation
does not significantly cool the nuggets: the thermal radiation from
the Boltzmann regime discussed in~\cite{Forbes:2008uf} dominates.

\section{Diffuse Galactic Emissions}
\label{sec:diffuse-1-20}\noindent
Previous discussions of quark (anti)nugget dark matter considered the
electrosphere in only two limiting cases: the inner high-density
ultrarelativistic regime and the outer low-density Boltzmann regime.
While this analysis was sufficient to qualitatively discuss the
different components of the spectrum, it did not allow for their
comparison in any level of detail.  In particular, comparing the
relative strengths of the 511~keV line emission (emitted entirely from
the nonrelativistic regime) with that of the MeV continuum, (emitted
from the relativistic regime) required introducing a phenomenological
parameter $\chi \approx 0.1$~\cite{Lawson:2007kp} to express the
relative rates of direct annihilation to positronium formation.  This
parameter is sensitive to the density profile at all scales.  Using
the detailed numerical solutions of the previous section we can
directly compute the relative intensities of the two emission
mechanisms.  Our calculation demonstrates that this value of $\chi$ is
supported by a purely microscopic calculation rooted firmly in $\QED$
and well-understood many-body physics. The agreement between the
calculated value of $\chi$ and the phenomenological value required to
fit the observations provides another important and nontrivial test
of our dark-matter proposal. In addition, using the same numerical
solutions of the previous section, we compute the spectrum in the few
MeV region which could not be calculated in~\cite{Lawson:2007kp} with
only the ultrarelativistic approximation for the density profile.

\subsection{Observations}
\subsubsection{The 511~keV Line}\noindent 
S\textsc{pi}/\INTEGRAL\ has detected a strong 511~keV signal from the Galactic
bulge~\cite{Weidenspointner:2006nu}.  A spectral analysis shows this to be
consistent with low-momentum $e^+e^-$ annihilation through a positronium
intermediate state: About one quarter of positronium decays emit two 511~keV
photons, while the remaining three-quarters will decay to a three photon state
producing a continuum below 511~keV.  Both emissions have been seen by
\SPI/\INTEGRAL\ with the predicted ratios.

The 511~keV line is strongly correlated with the Galactic centre with
roughly 80\% of the observed flux coming from a circle of half angle
6$^\circ$.  There also appears to be a small asymmetry in the
distribution oriented along the Galactic disk~\cite{2008Natur.451..159W}.

The measured flux from the Galactic bulge is found to be
$\d\Phi/\d\Omega \simeq
0.025$~photons~cm$^{-2}$~s$^{-1}$~sr$^{-1}$~\cite{Jean:2003ci}.  After
accounting for all known Galactic positronium sources the 511~keV line
seems too strong to be explained by standard astrophysical
processes. These processes, as we understand them, seem incapable of
producing a sufficient number of low-momentum positrons.  Several
previous attempts have been made to account for this positron
excess. Suggestions have included both modifications to the understood
spectra of astrophysical objects or positrons which arise as a final
state of some form of dark-matter annihilation.  At this time there is
no conclusive evidence for any of these proposals.

If we associate the observed 511~keV line with the annihilation of low momentum
positrons in the electrosphere of an antiquark nugget, these properties are
naturally explained. The strong peak at the Galactic centre and extension into
the disk must arise because the intensity follows the distribution
$\rho_{\textsc{v}}(\vect{r})\rho_{\textsc{dm}}(\vect{r})$ of visible and dark-matter
densities. This profile is unique to dark-matter models in which the observed
emission is due to matter--dark matter interactions and should be contrasted
with the smoother $\sim \rho_{\textsc{dm}}^2(\vect{r})$ profile for proposals based on
self-annihilating dark-matter particles, or the $\sim\rho_{\textsc{dm}}(\vect{r})$
profile for decaying dark-matter proposals. The distribution
$\rho_{\textsc{v}}(\vect{r})\rho_{\textsc{dm}}(\vect{r})$ obviously implies that the predicted
emission will be asymmetric, with extension into the disk from the Galactic
center as it tracks the visible matter.  There appears to be evidence for an
asymmetry of this form~\cite{2008Natur.451..159W}.  In our proposal, no
synchrotron emission will occur as the positrons are simply an integral part of
the nuggets rather than being produced at high energies with only the low-energy
components exposed for emission.  Contrast this with many other dark matter
based proposals where the relatively high-energy positrons produced from
decaying or annihilating dark-matter particles produce strong synchrotron
emission.  These emissions are typically in conflict with the strong
observational constraints~\cite{Beacom:2005qv}.

\subsubsection{Diffuse MeV scale emission}
\label{sec:diffuse-mev-scale}\noindent
The other component of the Galactic spectrum we discuss here is the diffuse
continuum emission in the 1 -- 30~MeV range observed by \COMPTEL\/\CGRO.  The
interpretation of the spectrum in this range is more complicated than that of
the 511~keV line as several different astrophysical processes contribute.

The conventional explanation of these diffuse emissions is that of
gamma rays produced by the scattering of cosmic rays off of the
interstellar medium, and while detailed studies of cosmic ray
processes provide a good fit to the observational data over a wide
energy range (from 20~MeV up to 100~GeV), the predicted spectrum falls
short of observations by roughly a factor of 2 in the 1 -- 20~MeV
range~\cite{Strong:2004de}.

Background subtraction is difficult, however, and especially obscures the
spatial distribution of the MeV excess.  However, the excess seems to be
confined to the inner Galaxy ($l=330^\circ$ -- $30^\circ, |b| =0^\circ$ --
$5^\circ$), \cite{Strong:2004de} with a negligible excess from outside of the
Galactic centre.

\subsubsection{Comparison}\noindent
As our model predicts both of these components to have a common
source, a comparison of these emissions provides a stringent tests of
the theory.  In particular, the morphologies, spectra, and relative
intensities of both emissions must be strongly related.

The inferred spatial distribution of these emissions is consistent: both are
concentrated in the core of the Galaxy. Unfortunately, this is not a very stringent
test due to the poor spatial resolutions of the present observations.  The
prediction remains firm: If the morphology of the observations can be improved,
then a full subtraction of known astrophysical sources should yield a diffuse
MeV continuum with spatial morphology identical to that of the 511~keV.

Present observations, however, do allow us to test the intensity and
spectrum predicted by the quark nugget dark-matter model.  The model
predicts the intensity to be proportional to the 511~keV flux with
calculable coefficient of proportionality.  Therefore, the \INTEGRAL\
data may be used to fix the total diffuse emission flux, removing the
uncertainties associated with the line of sight averaging which is the
same for both emissions.

The resulting spectrum and intensity were previously discussed
in~\cite{Zhitnitsky:2006tu} and~\cite{Lawson:2007kp}, but the 511~keV line
emission was estimated using the low-density Boltzmann approximation, while the
MeV emissions were estimated using the high-density ultrarelativistic
approximation.  The proportionality factor linking these two was treated as a
phenomenological parameter, $\chi$, which required a value of $\chi \approx
0.1$~\cite{Lawson:2007kp} in order to explain the observations.  Here, equipped
with a complete density profile, we calculate the value of this parameter from
first principles showing that it is indeed consistent with the observations,
providing yet another highly nontrivial verification of the quark antimatter
nugget dark-matter proposal.

\subsection{Annihilation rates}\noindent
As incident electrons enter the electrosphere of positrons, the dominant
annihilation process is through positronium formation with the subsequent
annihilation producing the 511~keV emission.  As the remaining electrons
penetrate more deeply, they will encounter a higher density of more energetic
positrons: Direct annihilation eventually becomes the dominant process. The
maximum photon energy, $\sim 10$~MeV, is determined by the Fermi energy of the
electrosphere at the deepest depth of electron penetration.  We stress that this
scale is not introduced in order to explain the \COMPTEL/\CGRO\ gamma-ray
excess: Our model necessarily produces a strong emission signature at precisely
this energy.

In the following section we integrate the rates for these two processes 
over the complete density profile, allowing us to predict the relative 
intensities and spectral properties of the
emissions.

\subsubsection{Positronium Formation}\noindent
In principle, we only need to know the $e^+e^-$ annihilation
cross section at all centre of mass momenta.  At high energies, this
is perturbative and one can use the standard \textsc{qed} result for
direct $e^+e^- \rightarrow 2\gamma$ emission.  At low energies,
however, one encounters a strong resonance due to the presence of a
positronium state the greatly enhances the emission.  This resonance
renders a perturbative treatment invalid and must be dealt with
specially, thus we consider these as two separate processes.  Our
presentation here will be abbreviated: details may be found in
\cite{Zhitnitsky:2006tu}.

While the exact positronium formation rate -- summed over all excited states --
is not well established, it is clear that the rate will fall off rapidly as the
centre of mass momentum moves away from resonance.  A simple estimate suggests
that for momenta $p > m\alpha$ the formation rate falls as $ \sim p^{-4}$.  To
estimate the rate, we thus make a cutoff at $q \approx m\alpha$ as the upper
limit for positronium formation: for large center of mass momenta we use the
perturbative direct-annihilation approximation.  The scale here is set by the
Bohr radius $a_b = 2(m\alpha)^{-1}$ for the positronium bound state: If a
low-momentum $e^+e^-$ pair pass within this distance, then the probability of
forming a bound state becomes large, with a natural cross section $\sigma_{Ps}
\sim \pi a_b^2$. The corresponding rate is
\begin{equation}
  \label{eq:gamma_Ps}
  \Gamma_{Ps} = \int_{p\lesssim q}\!\!\!\! v \sigma_{Ps}n(p)
  \frac{\d^3{p}}{(2\pi)^3}
  \sim \frac{4v}{m^2\alpha^2}\begin{cases}
    n(p) \pi & p_F \lesssim q\\
    \frac{q^3}{3\pi} & p_F \gtrsim q.
  \end{cases}
\end{equation}
where $n(p)$ is the momentum distribution of the positrons in the
electrosphere, and $v \sim \alpha$ is the incoming electron velocity.
The second expression represents the two limits (valid at low
temperature): a) of low density when the Fermi momentum $p_F\lesssim
q$ and all states participate, resulting in a factor of the total
density $n$, and b) of high density where the integral is saturated by
$q^3/3\pi^2 \approx (m \alpha)^3/3\pi^2$.

As discussed in Sec.~\ref{sec:nugg-charge-equil}, and in more detail in
Appendix~\ref{sec:debye-1}, electrons will be able to penetrate the charged
antimatter nuggets in spite of the strong electric field.  Initially the
electrons are bound in neutral atoms.  Once ionized, the density is sufficiently
high that the charge is efficiently screened with a Debye screening length
$\lambda_D$ that is much smaller than the typical de Broglie wavelength
$\lambda=\hbar/p$ of electrons.  Thus, the electric fields -- although quite
strong in the nugget's electrosphere -- will not appreciably effect the motion.
The binding of electrons in neutral atoms complicates the analysis slightly, but
the binding energy -- on the eV scale -- will not significantly alter the
qualitative nature of our estimates.  For a precision test of the emission
properties, this will need to be accounted for.  Here we include bands
comprising $\pm 10\%$ relative variation in the overall positronium annihilation
rate to show the sensitivity to this uncertainty.

\subsubsection{Direct-annihilation Rates}\noindent
While positronium formation is strongly favoured at low densities due
to its resonance nature, deep within the electrosphere the rapidly
growing density of states at large momentum values result in a
cross section characterized by the perturbative direct-annihilation
process. Conceptually one can imagine an incident Galactic
electron first moving through a Fermi gas of positrons with roughly
atomic density with a relatively large probability of annihilation
through positronium to 511~keV photons.  A small fraction survive to
penetrate to the inner high0density region where direct annihilation
dominates. The surviving electrons then annihilate with a high-energy
positron near to the inner quark-matter surface releasing high-energy
photons.  The spectrum of these annihilation events will be a broad
continuum with an upper cutoff at an energy scale set by the Fermi energy
of the positron gas at the maximum penetration depth of the electrons.
The spectral density for direct $e^+e^-$ annihilation at a given
chemical potential was calculated in~\cite{Lawson:2007kp}:
\begin{align}
  \label{eq:specden}
  \frac{\d{I(\omega,\mu)}}{\d{}\omega \d{t}} 
  &= \int \d{n_{p}(\mu)}\; v_{p}(\mu) \frac{\d\sigma
    (p,\omega)}{\d{\omega}}\\
  &= \int \frac{\d^3{p}}{(2\pi)^3}\frac{2}{1+\e^{(\mu-E_{p})/T}}
  \frac{p}{E_{p}}\frac{\d\sigma(p,\omega)}{\d{\omega}}  \nonumber \\
  \frac{\d\sigma(p,\omega)}{\d{\omega}} &=\frac{\pi\alpha^2}{mp^2} 
  \Biggl[ \frac{-(3m+E_{p})(m+E_{p}) }{(m+E_{p}-\omega)^2} -2 \\
  &+ \frac{ \frac{1}{\omega} (3m+E_{p})(m+E_{p})^2
      -(\frac{m}{\omega})^2(m+E_{p})^2}{(m+E_{p}-\omega)^2}\Biggr],\nonumber
\end{align}
where $E_{p} = \sqrt{p^2 + m^2}$ is the energy of the positron in the
rest frame of the incident (slow-moving) electron and $\omega$ is the
energy of the produced photons.  The annihilation rate at a given
density, $\Gamma_{\text{dir}}(n_{\mu})$, is obtained by integrating
over allowed final state photon momentum. This was previously done in
the $T \rightarrow$ 0 limit where the integral may be evaluated
analytically, at nonzero temperatures the rate must be evaluated
numerically.  

\begin{figure}[tb]
  \begin{center}
    \includegraphics[width=\columnwidth]{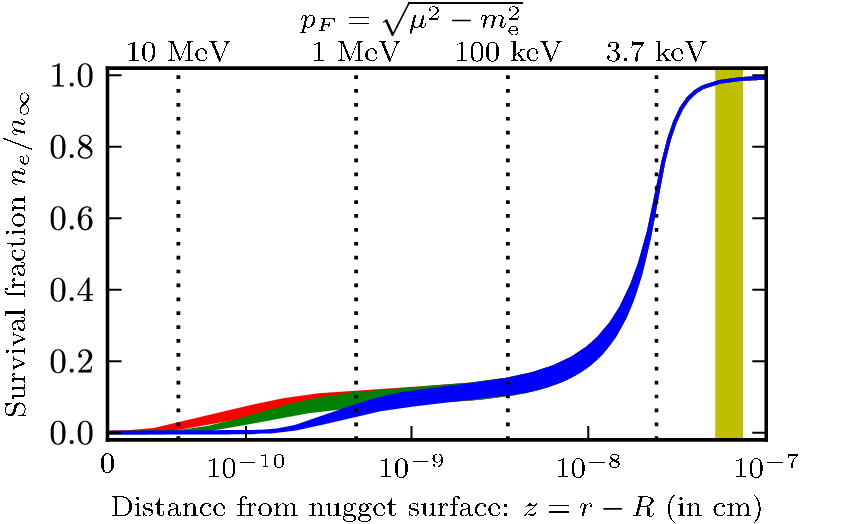}
    \caption{\label{fig:survival} Electron survival
      fraction, $n_{e}(r)/n_{\infty}$~(\ref{eq:survival}), of an
      incoming electron with velocities $v = 0.01c$ [leftmost light
      gray (red) band], $v = 0.005c$ [middle (green) band] and $v =
      0.001c$ [rightmost dark gray (blue) band] from left to right,
      respectively.  The thickness of the bands includes a $\pm 10\%$
      variation in the positronium annihilation
      rate~(\ref{eq:gamma_Ps}).  The local Fermi momentum $p_F$ is
      shown along the top and with vertical dotted lines including the
      cutoff scale $q\approx 3.7$ keV from~(\ref{eq:gamma_Ps}).  The
      yellow shaded Boltzmann region and the abscissa scaling are the
      same as in the top of Fig.~\ref{fig:antinugget_profiles}.  Note
      that annihilation happens well within the electrosphere, so the
      finite size of the nugget is irrelevant.}
  \end{center}
\end{figure}

\subsection{Spectrum and Branching Fraction}\noindent
To determine the full annihilation spectrum we first determine the
fraction of incident electrons that can penetrate to a given radius
$r$ in the electrosphere (see Fig.~\ref{fig:survival}.  We then
integrate the emissions over all regions.

Consider an incident beam of electrons with density $n_{\infty}$ and velocity
$v$.  As they enter the electrosphere of positrons, the electrons will
annihilate.  The survival fraction $n_{e}(r)/n_{\infty}$ will thus
decrease with a rate proportional to $\Gamma(r) = \Gamma_{Ps}(r) +
\Gamma_{\text{dir}}(r)$ which depends crucially on the local density profile
$n[\mu(r)]$ calculated in Sec.~\ref{sec:profiles}:
\begin{equation}
  \label{eq:decayDE}
  \frac{\d{n_{e}}(r)}{\d{t}} 
  = v^{-1}\frac{\d{n_{e}}(r)}{\d{r}} 
  = -\Gamma(r) v^{-1}n_{e}(r).
\end{equation}
Integrating (\ref{eq:decayDE}), we obtain the survival fraction:
\begin{equation}
  \label{eq:survival}
  \frac{n_{e}(r)}{n_{\infty}} 
  = \exp\left(-\int_{r}^{\infty} \d{r}\; v^{-1}\Gamma(r)\right).
\end{equation}
This is shown in Fig.~\ref{fig:survival}.  One can clearly see  that in
the outer electrosphere, positronium formation -- independent of $v$
-- dominates the annihilation.  Once the density is sufficiently high
($p_F \gtrsim 1$~MeV), the direct-annihilation process dominates,
introducing a dependence on the velocity $v$ of the incident particle.

The initial velocity $v\sim 10^{-3}c$ is determined by the local relative
velocity of the nuggets with the surrounding \ISM.  This will depend on the
temperature of the \ISM, but the positronium annihilation rate is insensitive to
this.  As we mentioned previously, most of the electrons in the interstellar
medium are bound in neutral atoms, either as neutral hydrogen HI, or in
molecular form H$_2$.  (The ionized hydrogen HII represents a very small mass
fraction of interstellar medium.)  These neutral atoms and molecules will have
no difficulty entering the electrosphere.

The remaining bound electrons that do not annihilate through positronium will
ionize once they reach denser regions, and will acquire a new velocity set by a
combination of the initial velocity $v \sim 10^{-3}c$ and the atomic velocity
$v\sim \alpha \sim 10^{-2}c$ imparted to the electrons as they are ionized from
the neutral atoms: The latter will typically dominate the velocity scale.  This
only occurs in sufficiently dense regions where the Debye screening discussed in
Appendix~\ref{sec:debye-1} becomes efficient.  Hence, the electric fields will
not significantly alter the motion of the electrons after ionization.  The
direct-annihilation process depends on the final velocity $v$; as the dominant
contribution comes from ionization, this will remain relatively insensitive to
the \ISM.  During ionization, some fraction (roughly half) of the electrons will
move away from the core, but a significant portion will travel with this
velocity $v$ toward the denser regions.

Two other features of Fig.~\ref{fig:survival} should be noted.  First is the
value of the survival fraction $\chi \sim 0.1$ at which direct annihilation
dominates [see Eq.~\ref{eq:branching}].  This is the value that was postulated
phenomenologically in~\cite{Lawson:2007kp} in order to explain the relative
intensities of the 511~keV and direct-annihilation spectra.  Here this value
results naturally from a microscopic calculation in a highly nontrivial manner
that depends on the structure of the density profile at all scales up to
$p_F=1$~MeV.  Second is the rapid increase in density near the core quickly
extinguishes any remaining electrons.  As such, even if the chemical potential
at the surface is larger -- $\mu_R \sim 100$~MeV or so -- virtually no Galactic
electrons penetrate deeply enough to annihilate at these energies. For
nonrelativistic electrons the maximum energy scale for emission is quite
generally set at the $\sim$ 20~MeV scale, depending slightly on $v$.

Having established the survival fraction as a function of height it is
now possible to work out the spectral density that will arise from the
$e^+e^-$ annihilations.  At a given height we can express the number
of annihilations through a particular channel as,
\begin{equation}
  \frac{\d{n}}{\d{r}} = n_{e}(r)v^{-1}\Gamma(r).
\end{equation}
Integrating this expression over all heights will then give the total
fraction of annihilation events proceeding via positronium $f_{511}$, and
via direct annihilation $f_{\text{MeV}}$.  (The numerical
values are given for $q=m\alpha$ and are quite insensitive to $v$.)
\begin{subequations}
  \begin{align}
    \label{eq:nPs}
    f_{511} &= \int_R^{\infty}\!\!\!\d{r}\;
    \frac{n_{e}(r)}{n_{\infty}}v^{-1}\Gamma_{Ps}(r) \approx 0.9, \\
    \label{eq:nMeV}
    f_{\text{MeV}} &= \int_R^{\infty} \d{r}\;
    \frac{n_{e}(r)}{n_{\infty}}v^{-1}\Gamma_{\text{dir}}(r)
    \approx 0.1.
  \end{align}
\end{subequations}
Note that in both cases, the overall normalization depends only on the
total rate of electron collisions with dark matter through
$vn_{\infty}$. As such, the ratio of 511~keV photons to MeV continuum
emission is independent of the relative densities (though it
will show some dependence on the local electron velocity
distribution).  Numerically, with a $\pm 10\%$ variation in the
positronium annihilation rate, we find
\begin{equation}
  \label{eq:branching}
  \chi = \frac{f_{\text{MeV}}}{f_{511}} \approx 0.05 - 0.2,
\end{equation}
which is quite insensitive to the velocity $v$.  This ratio was introduced as a
purely phenomenological parameter in~\cite{Lawson:2007kp} to explain the
observations.  Here we have calculated from purely microscopic considerations
that, for a wide range of nugget parameters, the required value $\chi \sim
0.1$~\cite{Lawson:2007kp} arises quite naturally.

We now have everything needed to compute the spectral density of the
MeV continuum.  The spectral density at a fixed chemical potential is
given by Eq.~\eqref{eq:specden}: this must now be averaged along the
trajectory of the incoming electron weighted by the survival
fraction~(\ref{eq:survival}) and the time spent in the given region of
the trajectory (set by the inverse velocity $v^{-1}$):
\begin{equation}
  \label{eq:finalspec}
  \frac{\d{I_{\text{total}}}}{\d{\omega}}
  = \int_R^{\infty}\d{r}\; 
  v^{-1}
  \frac{n_{e}(r)}{n_{\infty}}
  \frac{\d{I(\mu(r))}}{\d{\omega}\;\d{t}}.
\end{equation}

\begin{figure}[htbp]
  \begin{center}
    \includegraphics[width=\columnwidth]{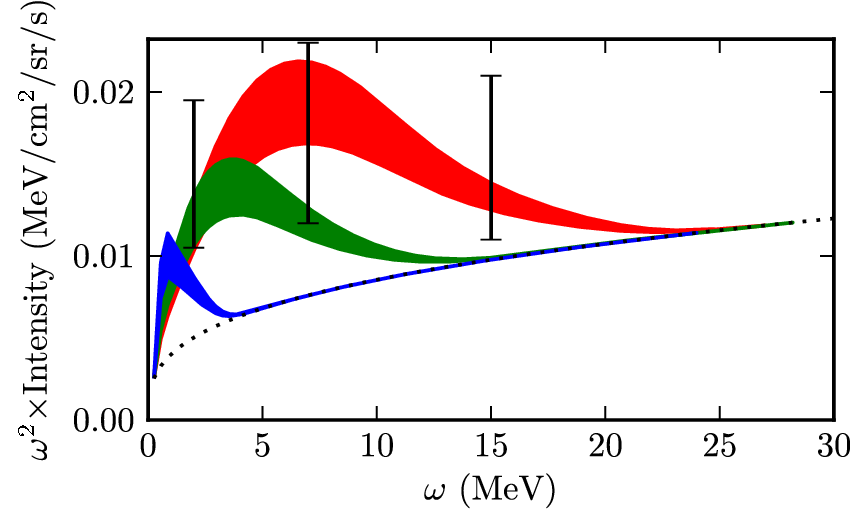}
    \caption{\label{fig:specplot} Spectral density (scaled by $\omega^2$ to
      compare with \cite{Strong:2004de} of photons emitted by an electron
      annihilating on antiquark nuggets with incoming velocities $v = 0.01c$
      [uppermost (red) band], $v = 0.005c$ [middle (green) band] and $v =
      0.001c$ [lowest (blue) band to the lower left] from right to left,
      respectively, including the cosmic ray background determined in
      \cite{Strong:2004de} (dotted line).  The thickness of the bands includes a
      $\pm 10\%$ variation in the positronium annihilation
      rate~(\ref{eq:gamma_Ps}).  The overall normalization is fixed to the
      observationally unrelated 511 keV line as discussed
      below~(\ref{eq:511norm}).  The three error bars are the \textsc{comptel}
      data points.  Note: This spectrum should still be interpreted as a
      qualitative effect - a detailed calculation of the ionization and hence
      distribution of the velocity $v$ must be performed to yield a quantitative
      prediction.  The general structure and magnitude, however, can be trusted
      as these depend on the overall density profile which we have carefully
      modeled.  (Compare with Fig.~\ref{fig:specplot_UR}, for example, which
      uses only the ultrarelativistic density profile: The resulting intensity
      is 2 orders of magnitude too large.) }
  \end{center}
\end{figure}

The resulting spectrum is shown in Fig.~\ref{fig:specplot} and is
sensitive to both the incoming velocity $v$ (which will depend on the
local environment of the antimatter nugget) and the overall
normalization of the positronium annihilation
rate~(\ref{eq:gamma_Ps}).  (The latter is fixed in principle, but
requires a difficult in-medium calculation to determine precisely.)
These two parameters are rather orthogonal. The velocity $v$
determines the maximum depth of penetration, and hence the maximum
energy of the emitted photons (as set by the highest chemical
potential at the annihilation point): If the electron velocity is
relatively low ($v<100$~km/s $\approx 0.0003c$), almost all
annihilations happen immediately and the MeV continuum will fall
rapidly beyond 5~MeV. As the velocity increases the electrons are able
to penetrate deeper toward the quark surface and annihilate with
larger energies.  In contrast, the details of the positronium
annihilation do not alter the spectral shape, but do alter the overall
normalization. 

As already mentioned earlier, the parameter $\chi\sim 0.1$ is calculated here
from purely microscopic physics.  Therefore, it provides highly nontrivial
verification of the entire proposal.  Also, the profile function from the
previous section is computed at all scales, allowing us to calculate the photon
spectrum down to, and below, the electron mass. This calculation could not be
performed in the previous analysis~\cite{Lawson:2007kp} with only the
ultrarelativistic expression.

We stress here that the result shown in Fig.~\ref{fig:specplot} exhibits several
nontrivial features arising from very general characteristics of the proposed
emission model.  The wide range of positron energies within the electrosphere
implies a broad emission profile with a width $\Delta \omega \simeq 5$ -- $10$
MeV. The initial growth of emission strength with $\omega$ is due to the
increasing density of states as a function of depth.  Above $\sim 10$ MeV, the
emission becomes suppressed by the inability of Galactic electrons to penetrate
to depths where the positrons have this energy.  While the exact details may
vary with parameters, such as the local electron velocity and precise rate of
positronium formation, the general spectral features are inescapable
consequences of our model, allowing it to be tested by future, more precise, observations.

\subsection{Normalization to the 511~keV line}\noindent
To obtain the observed spectrum, one needs to
average~(\ref{eq:finalspec}) along the line of site over the varying
matter and dark-matter density and velocity distributions.  Neither of
these is known very well, so to check whether or not the prediction is
significant, we fix the average rate of electron annihilation with the
observed 511~keV line which our model predicts to be produced by
the same process.  As the intensity of the 511~keV line (resulting
from the two-photon decay of positronium in the $^1$S$_0$ state) has
been measured by \textsc{spi}/\textsc{integral} along the line of
sight toward the core of the Galaxy (see~\cite{Weidenspointner:2008zl}
for a review), we can use this to fix the normalization of the MeV
spectrum along the same line of sight.  The total intensity is given
in terms of Eq.~\eqref{eq:nPs}:
\begin{equation}
  \label{eq:511norm}
  \frac{\d{\Phi}}{\d{\Omega}} = \frac{C}{4} f_{511} 
  \sim \;0.025\; \frac{\text{photons}}{\text{cm}^{2}\text{~s}\text{~sr}}\;,
\end{equation}
where the factor of $4$ accounts for the three-quarters of the
annihilation events that decay via the $^{3}$S$_{1}$ channel (also
measured, but not included in the line emission).  This fixes the
normalization constant $C\approx
0.11$~events~cm$^{-2}$~s$^{-1}$~sr$^{-1}$.  The
predicted contribution to the MeV continuum must have the same
morphology and, thus, an integrated intensity along the same line of
sight must have the same normalization factor.  This normalization has
been used in Fig.~\ref{fig:specplot} to compare the predicted
spectrum with the unaccounted for excess emission detected by
\textsc{comptel} (\cite{Strong:2004de}).

The exact shape of the spectrum will be an average of the components
shown over the velocity distribution of the incident matter.  Our
process of normalization is unable to remove this ambiguity because
the predicted 511 keV spectral properties are insensitive to this.  It
is evident, however, that MeV emission from dark-matter nuggets could
easily provide a substantial contribution to the observed MeV excess.
Note also, that the excess emission can extend only to 20 MeV or so.
This is completely consistent with the more sophisticated background
estimates discussed in~\cite{Porter:2008fk} which can fit almost all
aspects of the observed spectrum \emph{except} the excess between 1
and 30 MeV predicted by our proposal.

\section{Conclusion}\noindent
Solving the relativistic Thomas-Fermi equations, we determined the
charge and structure of the positron electrosphere of quark antimatter
nuggets that we postulate could comprise the missing dark-matter in
our Universe.  We found the structure of the electrosphere to be
insensitive to the size of the nuggets, as long as they are large
enough to be consistent with current terrestrial based detector
limits, and hence, can make unambiguous predictions about electron
annihilation processes.  

To test the dark-matter postulate further, we used the structure of
this electrosphere to calculate the annihilation spectrum for incident
electronic matter.  The model predicts two distinct components: a 511
keV emission line from decay through a positronium intermediate and an
MeV continuum emission from direct-annihilation processes deep within
the electrosphere.  By fixing the general normalization to the
measured 511 keV line intensity seen from the core of the Galaxy, our
model makes a definite prediction about the intensity and spectrum of
the MeV continuum spectrum without any additional adjustable model
parameters: Our predictions are based on well-established physics.

As discussed in~\cite{Lingenfelter:2009fk}, a difficulty with most other
dark-matter explanations for the 511 keV emission is to explain the large
$\sim$100\% observed positronium fraction -- positrons produced in hot regions
of the Galaxy would produce a much smaller fraction.  Our model naturally
predicts this observed ratio everywhere.

\textit{A priori}, there is no reason to expect that the predicted MeV spectrum
should correspond to observations: typically two uncorrelated emissions are
separated by many orders of magnitude.  We find that the phenomenological
parameter $\chi$~(\ref{eq:branching}) required to explain the relative
normalization of MeV emissions arises naturally from our microscopic
calculation.  This is highly nontrivial because it requires a delicate balance
between the two annihilation processes from the semirelativistic region of
densities that is sensitive to the semirelativistic self-consistent structure of
the electrosphere outside the range of validity of the analytic
ultrarelativistic and nonrelativistic regimes.  (See
Fig.~\ref{fig:antinugget_profiles} and~\ref{fig:survival}).

If the predicted emission were several orders of magnitude too large, the
observations would have ruled out our proposal.  If the predicted emissions were
too small, the proposal would not have been ruled out, but would have been much
less interesting.  Instead, we are left with the intriguing possibility that
both the 511 keV spectrum and much of the MeV continuum emission arise from the
annihilation of electrons on dark-antimatter nuggets.  While not a smoking gun
-- at least until the density and velocity distributions of matter and
dark-matter are much better understood -- this provides another highly
nontrivial test of the proposal that, \textit{a priori}, could have ruled it
out.

Both the formal calculations and the resulting structure presented here --
spanning density regimes from ultrarelativistic to nonrelativistic -- are
similar to those relevant to electrospheres surrounding strange-quark stars
should they exist. Therefore, our results may prove useful for studying quark
star physics.  In particular, problems such as bremsstrahlung emission from
quark stars originally analyzed in~\cite{Jaikumar:2004rp} (and corrected
in~\cite{Caron:2009zi}) that uses only ultrarelativistic profile functions.  The
results of this work can be used to generalize the corresponding analysis for
the entire range of allowed temperatures and chemical potentials. Another
problem which can be analyzed using the results of the present work is the study
of the emission of energetic electrons produced from the interior of quark
stars. As advocated in~\cite{Charbonneau:2009ax}, these electrons may be
responsible for neutron star kicks, helical and toroidal magnetic fields, and
other important properties that are observed in a number of pulsars, but are
presently unexplained.

Finally, we would like to emphasize that this mechanism demonstrates that dark
matter may arise from \emph{within} the standard model at the \QCD\
scale,\footnote{The axion is another dark-matter candidate arising from the
  \QCD\ scale, but with fewer observational consequences.} and
that exotic new physics is not required.  Indeed, this is naturally suggested by
the ``cosmic coincidence'' of almost equal amounts of dark and visible
contributions to the total density $\Omega_{\text{tot}} = 1.011(12)$ of our
Universe~\cite{Amsler:2008zz}:
\begin{equation*}
  \Omega_{\text{dark-energy}}
  : \Omega_{\text{dark-matter}}
  : \Omega_{\text{visible}} \approx 17:5:1.
\end{equation*}
The dominant baryon contribution to the visible portion $\Omega_{\text{visible}}
\approx \Omega_{B}$ has an obvious relation to \QCD\ through the nucleon mass
$m_{\textsc{n}} \propto \Lambda_{\QCD}$ (the actual quark masses arising from
the Higgs mechanism contribute only a small fraction to $m_{\textsc{n}}$).
Thus, a \QCD\ origin for the dark components would provide a natural solution to
the extraordinary ``fine-tuning'' problem typically required by exotic
high-energy physics proposals.  Our proposal here solves the matter portion of
this coincidence.  For a proposal addressing the energy coincidence we refer the
reader to \cite{Urban:2009vy,*Urban:2009yg} and references therein.\footnote{The
  idea concerns the anomaly that solves the famous axial $U(1)_{A}$ problem,
  giving rise to an $\eta'$ mass that remains finite, even in the chiral limit.
  Under some plausible and testable assumptions about the topology of our
  Universe, the anomaly demands that the cosmological vacuum energy depend on
  the Hubble constant $H$ and \QCD\ parameters as $\rho_{\textsc{de}} \sim H
  m_q\langle\bar{q}q\rangle /m_{\eta'} \approx (4 \times 10^{-3}\text{ eV})^4$
  -- tantalisingly close to the value $\rho_{\textsc{de}} =
  [1.8(1)\times 10^{-3}\text{ eV}]^4$ observed today~\cite{Amsler:2008zz}.}

\begin{acknowledgments}
  M.M.F. would like to thank George Bertsch, Aurel Bulgac, Sanjay Reddy, and
  Rishi Sharma for useful discussions, and was supported by the LDRD program at
  Los Alamos National Laboratory, and the U.S. Department of Energy under grant
  number DE-FG02-00ER41132.  K.L and A.R.Z. were supported in part by the
  Natural Sciences and Engineering Research Council of Canada.
\end{acknowledgments}

\paragraph*{Note in proof:} While this paper was in preparation, a
review~\cite{Prantzos:2010} was submitted which mentions the mechanism discussed
here, but dismisses it based on the arguments of~\cite{Cumberbatch:2006bj}.
This work makes it clear that the assumptions used in~\cite{Cumberbatch:2006bj}
are invalid (see footnote~\ref{foot:cumberbatch}) and addresses the issues
raised therein.  The arguments of~\cite{Cumberbatch:2006bj} were previously
addressed in~\cite{Forbes:2008uf} (at the end of appendix 1) where it was
emphasized that~\cite{Cumberbatch:2006bj} incorrectly applies a relativistic
formula to the non-relativistic regime.

\appendix
\section{Density Functional Theory}
\label{sec:dens-funct-theory}\noindent
We start with a Density Functional Theory (\DFT) formulated in terms
of the thermodynamic potential:
\begin{subequations}
  \begin{multline}
    \label{eq:DFT}
    \Omega = \sum_{i}2f_{i}\int\d^{3}\vect{r}\;
    \psi^{\dagger}_{i}(\vect{r})(\epsilon_{-i\hbar\vect{\nabla}}-\mu)\psi_{i}(\vect{r}) + \\
    + \int\d^{3}\vect{r} V_{\text{ext}}(\vect{r}) n(\vect{r}) 
    + \int\d^{3}\vect{r}\;\varepsilon_{\text{xc}}\Bigl[n(\vect{r})\Bigr]n(\vect{r}) +\\
    + \frac{4\pi e^2}{2}\int\d^{3}{\vect{r}}\;\d^{3}{\vect{r}'}
      \;\frac{n(\vect{r})n(\vect{r}')}{\norm{\vect{r}-\vect{r}'}}
    - T\sum_{i} f_{i}\ln f_{i},
  \end{multline}
  where $\epsilon_{p}=\sqrt{p^2 + m^2}$ is the relativistic energy of the
  electrons, $\varepsilon_{\text{xc}}(n)$ is the exchange energy.  The density is
  \begin{equation}
    \label{eq:Fermi-Dirac}
    n(\vect{r}) = 2\sum_{i}f_{i}\psi_{i}^{\dagger}(\vect{r})\psi_{i}(\vect{r}),
  \end{equation}
  where we have explicitly included the spin degeneracy, and used
  relativistic units where $\hbar = c = 1$ and $e^2 = \alpha$.

  We vary the potential with respect to the occupation numbers $f_{i}$
  and the wave functions $\psi_{i}$ subject to the constraints that the
  wave functions be normalized,
  \begin{equation}
    \label{eq:Normalization}
    \int \d^{3}\vect{r}\; \psi_{i}^{\dagger}(\vect{r})\psi_{j}(\vect{r}) = \delta_{ij},
  \end{equation}
  and that the occupation numbers satisfy Fermi-Dirac
  statistics.\footnote{This is most easily realized by introducing the
    single-body density matrix $\mat{\rho} = \mat{1} -
    \mat{C}\mat{\rho}^{T}\mat{C}$ where $\mat{C}$ is the
    charge-conjugation matrix.}
\end{subequations}
This yields the Kohn-Sham equations for the electronic wave functions:
\begin{align*}
  \Bigl[\epsilon_{(-i\hbar \vect{\nabla})} - \mu + V_{\text{eff}}(\vect{r})\Bigr]
  \psi_{i}(\vect{r}) &= E_{i}\psi_{i}(\vect{r}),\\
  f_{i} &= \frac{1}{1+\e^{E_{i}/T}}.
\end{align*}
where (we take $e$ to be positive here)
\begin{equation*}
  V_{\text{eff}}(\vect{r}) = e\phi(\vect{r}) + V_{\text{ext}}(\vect{r}) + 
  \varepsilon_{\text{xc}}[n(\vect{r})] + n(\vect{r})\varepsilon_{\text{xc}}'[n(\vect{r})]
\end{equation*}
and
\begin{equation*}
  \phi(\vect{r}) = 4\pi e\int\d^{3}\vect{r}'\frac{n(\vect{r} +
    \vect{r}')}{\norm{\vect{r}}}
\end{equation*}
is the electrostatic potential obeying Poisson's equation,
\begin{equation*}
  \nabla^2 \phi(\vect{r}) = -4\pi e n(\vect{r}).
\end{equation*}

\subsection{Thomas-Fermi Approximation.}\noindent
\begin{subequations}
  If the effective potential $V_{\text{eff}}(\vect{r})$ varies sufficiently
  slowly, then it is a good approximation to replace it
  locally with a constant potential.  The Kohn-Sham equations thus
  become diagonal in momentum space, $i \equiv k$, $\psi_{k} \propto
  \e^{ikr}$ and may be explicitly solved:
  \begin{gather*}
    E_{\vect{k}}(\vect{r}) = \epsilon_{\vect{k}} - \mu + V_{\text{eff}}(\vect{r}),\\
    n(\vect{r}) = 2\int\frac{\d^{3}\vect{k}}{(2\pi)^3}\left(
      \frac{1}{1+\e^{E_{\vect{k}}(\vect{r})/T}} -
      \frac{1}{1+\e^{-E_{\vect{k}}(\vect{r})/T}} \right),
  \end{gather*}
  where we have also explicitly included the contribution from the
  antiparticles.  We piece these homogeneous solutions together at
  each point $\vect{r}$ and find the self-consistent solution that satisfies
  Poisson's equation.  This approximation is a relativistic
  generalization of the Thomas-Fermi approximation.
\end{subequations}

In principle, the \DFT\ method is exact~\cite{HK:1964}, however, the
correct form for $\varepsilon_{\text{xc}}$ is not known and could be
extremely complicated.  Various successful approximations exist but
for our purposes we may simply neglect this.  The weak
electromagnetic coupling constant $\alpha \sim 1/137$ and Pauli
exclusion principle keep the electrons sufficiently dilute that the
many-body correlation effects can be neglected until the density is
high in the sense that $\alpha^2 n \sim \alpha^2 \mu^3/3\pi^3 \gtrsim
m^3$, which corresponds to $\mu \gtrsim 50$~MeV.

Near the nugget core, many-body correlations may become quantitatively
important.  For example, the effective mass of the electrons in the gas
is increased by about $20\%$ when $\mu \approx 25$~MeV and doubles
when $\mu \approx 100$~MeV (see, for example,~\cite{Braaten:1992rz}).
These effects, however, will not change the qualitative structure, and
can be quite easily taken into account if higher accuracy is required.
We also note that, formally, the Thomas-Fermi approximation is only
valid for sufficiently high densities.  However, it gives correct
energies within factors of order unity for small
nuclei~\cite{LL3:1977} and is known to work substantially better for
large nuclei~\cite{LS:1973}.  As long as we do not attempt to use it
in the extremely low-density tails, it should give an accurate
description.

\subsection{Analytic Solutions}
\label{sec:analytic-solutions}\noindent
There are several analytic solutions available if we consider the
one-dimensional approximation, neglecting the curvature term
$2\mu'(r)/r$ in~(\ref{eq:RadialLaplacian}), which is valid close to
the nugget where $z = r - R \ll R$, the distance from the nugget
core, is less than the radius of the nugget.

\subsubsection{Ultrarelativistic Regime}
\label{sec:ultrarelat-regime}\noindent
The first is the ultrarelativistic approximation where $\mu$ and/or
$T$ are much larger than $m$ and the limit $m\rightarrow 0$ can be
taken.  In this case, we may explicitly evaluate (see
also~\cite{Kettner:1994zs, Cheng:2003le, Usov:2004kj}),\kern-1pt
\footnote{In this limit, then integrals have a closed form:
  \begin{multline}
    \label{eq:n_UR}
    n[\mu,T]_{m\ll \mu,T} =\\
    = \frac{1}{\pi^2}\int_{0}^{\infty}\d{p}\;
    \left[
      \frac{p^2}{1+\e^{(p-\mu)/T}}
      -
      \frac{p^2}{1+\e^{(p+\mu)/T}}
    \right] = \\
    = \frac{T^3}{\pi^2}\int_{0}^{\infty}\d{x}\;
    \left[
      \frac{x^2}{1+\e^{x-\mu/T}}
      -
      \frac{x^2}{1+\e^{x+\mu/T}}
    \right] = \\
    =\frac{-T^3}{\pi^2}\Gamma(3)\left[
      \Li_{3}(-\e^{\mu/T})
      -
      \Li_{3}(-\e^{-\mu/T})
    \right]
  \end{multline}
  where $\Li_{s}(z)$ is the Polylogarithm
  \begin{equation}
    \Li_{s}(z) = \sum_{k=1}^{\infty}\frac{z^{k}}{k^s}.
  \end{equation}}
\begin{equation}
  \label{eq:n_mu_UR}
  n[\mu,T]_{m\ll \mu,T} = \frac{\mu^3}{3\pi^2} + \frac{\mu T^2}{3}.
\end{equation}
In the domain-wall approximation, this admits an exact
solution with a typical length scale of $z_{\textsc{ur}}$:
\begin{gather}
  \mu(z,T) = \frac{T\pi\sqrt{2}}
  {\sinh\left[2T\sqrt{\frac{\alpha\pi}{3}}(z+z_{T})\right]},\\
  z_{\textsc{ur}} = \frac{1}{2T}\sqrt{\frac{3}{\alpha\pi}}
  \sinh^{-1}\left(\frac{T\pi\sqrt{2}}{\mu_{R}}\right).
\end{gather}
In our case, $T\sim 1\text{ eV } \ll m$ so we can also take the
$T\rightarrow 0$ limit to obtain~(\ref{eq:UR}):
\begin{align}
  \label{eq:n_z_UR}
  n_{\textsc{ur}}(z) &\approx 
  \frac{\mu_{R}^3}{3\pi^2(1 + z/z_{\textsc{ur}})^3}, &
  z_{\textsc{ur}} &\approx \mu_{R}^{-1}\sqrt{\frac{3\pi}{2\alpha}}.
\end{align}
This solution persists until $\mu \approx m$, which occurs at a
distance
\begin{equation}
  z_{\textrm{B}} \approx z_{\textsc{ur}}\left(\frac{\mu_{R}}{m} - 1\right)
  \approx m^{-1}\sqrt{\frac{3\pi}{2\alpha}}.
\end{equation}

\subsubsection{Boltzmann Regime}\noindent
Once the chemical potential is small enough that $\e^{\mu/T} \ll
\e^{m/T}$, we may neglect the degeneracy in the system, and write 
\begin{equation}
  \label{eq:n_mu_Boltzman}
  \left.n[\mu]\right|_{\exp(\mu/T) \ll \exp(m/T)} \approx n_{0}\e^{\mu/T}.
\end{equation}
This occurs for a density of about
\begin{equation}
  n_{\textrm{B}} \approx \left(\frac{mT}{2\pi}\right)^{3/2}
\end{equation}
and lower.  If the one-dimensional approximation is still valid, then
another analytic solution may be found:
\begin{align}
  n(z) &= \frac{n_{\textrm{B}}}
  {\left(1+\frac{z - z_{\textrm{B}}}{z_{\alpha}}\right)^2}, &
  z_{\alpha} &= \sqrt{\frac{T}{2\pi\alpha n_{\textrm{B}}}}
\end{align}
from which we obtained~(\ref{eq:Boltzmann}).  The shift $z_{\textrm{B}}$ must
be determined from the numerical profile at the point where $n(z_{\textrm{B}}) =
n_{\textrm{B}}$.  Note that $z_{\alpha} \ll z_{\textrm{B}}$, so this regime is valid and
persists until $z \sim 0.1 R$, at which point the one-dimensional
approximation breaks down.  It is in this Boltzmann regime where most
of the important radiative processes take place~\cite{Forbes:2008uf}.

\subsection{Numerical Solutions}
\label{sec:numerical-solutions}\noindent
The main technical challenge in finding the numerical solution is to
deal effectively with the large range of scales: $T \sim 1\text{ eV}
\ll m \sim 500\text{ keV} \ll \mu_R \sim 25\text{ MeV}$.  For example,
the density distribution~(\ref{eq:n_mu}) is prone to round-off error,
but the integral can easily be rearranged to give a form that is
manifestly positive:

\begin{equation}
  \label{eq:n_mu_good}
  n[\mu] = \frac{1}{\pi^2}
  \int_{0}^{\infty}\!\!\!\!\! \d{p}\;\frac{p^2 \sinh\left(\frac{\mu}{T}\right)}
  {\cosh\left(\frac{\sqrt{p^2 + m^2}}{T}\right) 
    + \cosh\left(\frac{\mu}{T}\right)}.
\end{equation}
The derivative may also be safely computed.  Let
\begin{align}
  A &= \frac{\sqrt{p^2 + m^2}}{T}, &
  B &= \frac{\mu}{T},
\end{align}
to simplify the expressions.  These can each be computed without any
round-off error.  The first derivative presents no further difficulties:
\begin{equation}
  \label{eq:dn_mu}
  \dot{n}[\mu] = \frac{1}{T\pi^2}
  \int_{0}^{\infty}\!\!\!\!\! \d{p}\;p^2
  \frac{1+\cosh{A}\cosh{B}}
  {\left(\cosh{A} + \cosh{B}\right)^2}.
\end{equation}
The differential equation is numerically simplified if we change
variables to logarithmic quantities.  We would also like to capture
the relevant physical characteristics of the solution, known from the
asymptotic regimes.  Close to the nugget, we have $\mu =
\mu_{R}/(1+z/z_{0})$, we introduce an abscissa logarithmic in the denominator
\begin{equation}
  a = \ln\left(1 + \frac{z}{z_{0}}\right)
   = \ln\left(1 + \frac{r-R}{z_{0}}\right).
\end{equation}
The dependent variable should be logarithmic in the chemical
potential, so we introduce $b = -\ln(\mu/\mu_{0})$.
We thus introduce the following change of variables:
\begin{subequations}
  \begin{align}
    r &= R + z_{0}(\e^{a} - 1),&
    \mu &= \mu_{0} \e^{-b},\\
    a &= \ln\left(1 + \frac{r - R}{z_{0}}\right), &
    b &= - \ln\frac{\mu}{\mu_{0}}.
  \end{align}
\end{subequations}
With the appropriate choice of scales $z_{0}$ describing the typical
length scale at the wall $r=R$ and $\mu_{0} \sim \mu_{R}$, these form
quite a smooth parametrization.  The resulting system is
\begin{multline}
  \ddot{b}(a) = \dot{b}(a)\left(\dot{b}(a) + 
    \frac{R-z_{0} -z_{0}\e^{a}}{R-z_{0} + z_{0}\e^{a}}\right) +\\
  -4\pi\alpha z_{0}^{2}\e^{2a} \frac{n[\mu]}{\mu}.
\end{multline}

\begin{figure}[htbp]
  \begin{center}
    \includegraphics[width=\columnwidth]{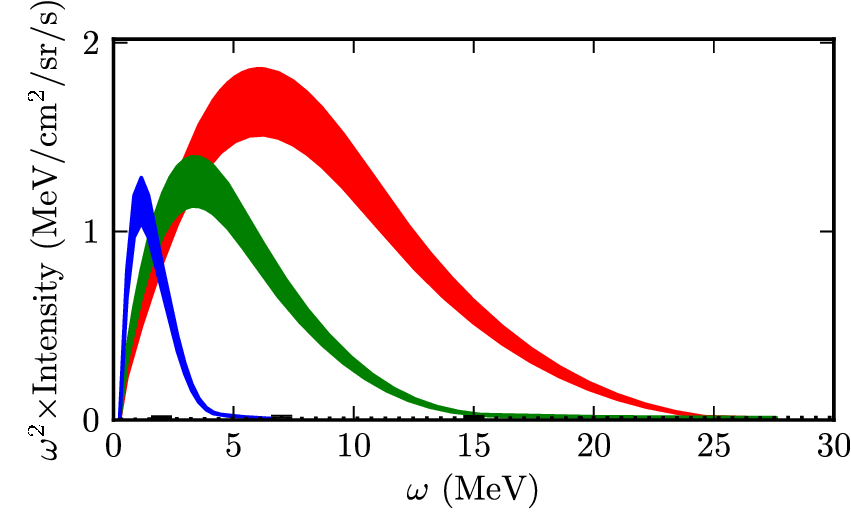}
    \caption{\label{fig:specplot_UR} Incorrect spectral density obtained by
      using a purely ultrarelativistic approximation for the
      profile~(\ref{eq:n_z_UR}).  Note that this is two orders of magnitude
      larger than the spectrum obtained from the full numerical solution
      (Fig.~\ref{fig:specplot}) and, if correct, would have easily ruled out
      our proposal.  This serves to demonstrate the highly nontrivial nature of
      the predicted spectrum shown in Fig.~\ref{fig:specplot} that is
      consistent with the observations~\cite{Strong:2004de}.}
  \end{center}
\end{figure}

It is imperative to include a full numerical solution to the profile
in order to obtain the proper emission spectrum.  Using only the
ultrarelativistic approximation~(\ref{eq:n_z_UR}) produces a spectrum
(Fig.~\ref{fig:specplot_UR}) 2 orders of magnitude too large, in
direct contradiction with the observations~\cite{Strong:2004de}.  The
actual prediction depends sensitively on a subtle -- but completely
model-independent -- balance between the ultrarelativistic,
relativistic, and nonrelativistic regimes.  The consistency between
the predicted spectrum shown in Fig~\ref{fig:specplot} and the
observations is a highly nontrivial test of the theory.

\section{Debye Screening In the Electrosphere}
\label{sec:debye-1}\noindent
Here we briefly discuss the plasma properties inside of the nugget's
electrosphere. The main point is that the Debye screening length
$\lambda_D$ is much smaller than the typical de Broglie wavelength
$\lambda=\hbar/p$ of electrons.  Thus, the electric fields -- although
quite strong in the nugget's electrosphere -- will not appreciably
effect the motion of electrons within the electrosphere as the charge
is almost completely screened on a scale $\sim \lambda_D$.

This screening effect was completely neglected in~\cite{Cumberbatch:2006bj},
which led the authors to erroneously conclude that all electrons will be
repelled before direct annihilation can proceed.  The overall charge (see
Table~\ref{tab:Antinugget}) will repel incident electrons at long distances as
discussed in Sec.~\ref{sec:nugg-charge-equil}, but neutral hydrogen will easily
penetrate the electrosphere, at which point the screening becomes effective,
allowing the electrons to penetrate deeply and annihilate as discussed in
Sec.~\ref{sec:diffuse-1-20}.

To estimate the screening of an electron, we solve the Poisson
equation with a $\delta^3(\vect{r}-\vect{r}_0)$ function describing the electron at
position $\vect{r}_0$~(see also \ref{eq:Poisson}),
\begin{equation}
  \label{eq:Poisson_1}
  \nabla^{2}\phi(\vect{r}) = 
  - 4\pi e \left[n(\vect{r}) - \delta^3 (\vect{r}-\vect{r}_0)\right]
\end{equation}
where $\phi(\vect{r})$ is the electrostatic potential and $n(\vect{r})$ is the
density of positrons.  As before, we exchange $\phi$ for $\mu$, the
effective chemical potential~(\ref{eq:mu}).  Now let $n_{0}(\vect{r})$ and
$\mu_{0}(\vect{r})$ be the solutions discussed in Sec.~\ref{sec:profiles}
without the potential $\delta^3(\vect{r} - \vect{r}_0)$.  We may then describe the
screening cloud by $\mu(\vect{r}) = \mu_{0}(\vect{r}) + \delta_{\mu}(\vect{r})$ and
$n(\vect{r}) = n_{0}(\vect{r}) + \delta_{n}(\vect{r})$ where
\begin{multline}
  \label{eq:Poisson_3}
  \nabla^{2}\delta_{\mu}(\vect{r}) 
  = 4\pi e^2 \left[
    \delta_{n}(\vect{r}) - \delta^3(\vect{r}-\vect{r}_{0})
  \right]\\
  \approx 
  4\pi \alpha \left[
    \pdiff{n[\mu]}{\mu}\delta_{\mu}(\vect{r}) - \delta^3(\vect{r}-\vect{r}_{0})
  \right]
\end{multline}
where $n[\mu]$ is given by~(\ref{eq:n_mu}).  If the density is
sufficiently large compared to the screening cloud deviations, we may
take $\partial{n}/\partial{\mu}$ to be a constant, in which case we
may solve~(\ref{eq:Poisson_3}) analytically with the boundary
conditions:
\begin{align}
  \lim_{\vect{r}\rightarrow \vect{r}_{0}}\delta_\mu(\vect{r}) 
  &= \frac{4\pi\alpha}{\norm{\vect{r} - \vect{r}_0}}, 
  & \lim_{\vect{r}\rightarrow \infty} \delta_\mu(\vect{r}) &= 0.
\end{align}
This gives the standard Debye screening solution
\begin{equation}
  \delta_{\mu}(\vect{r}) =
  \frac{-4\pi\alpha}{\norm{\vect{r}-\vect{r}_{0}}}
  \exp\left(-\frac{\norm{\vect{r}-\vect{r}_{0}}}{\lambda_{D}}\right)
\end{equation}
where the Debye screening length is
\begin{equation}
  \lambda_{D}^{-2}  = 4\pi\alpha \pdiff{n[\mu]}{\mu}.
\end{equation}
On distances larger than $\lambda_{D}$, the charge of the electron is
effectively screened.  In particular, we can neglect the influence of
the external electric field on motion of the electron if $\lambda_{D}$
is small compared to the de~Broglie wavelength $\lambda=\hbar/p$ of
the electrons:
\begin{equation}
  \label{eq:ratio-1}
  1 \ll \frac{\lambda}{\lambda_D}\sim
  \frac{1}{m v}\sqrt{4\pi\alpha \pdiff{n[\mu]}{\mu}}.
\end{equation}
In the ultrarelativistic limit~(\ref{eq:n_mu_UR}), one has
$\partial{n}/\partial{\mu} \approx \mu^2/\pi^2 + T^2/3$ whereas in the
Boltzmann limit~(\ref{eq:n_mu_Boltzman}), one has
$\partial{n}/\partial{\mu} \approx n_{0}/T$.

The relevant electron velocity scale in our problem is $v\sim
10^{-3}c$ ($T \sim 1$~eV).  This is the typical scale for electrons
ionized from neutral hydrogen.  For these electrons, one immediately
sees that the screening becomes significant in the Boltzmann regime
once the density is larger than
\begin{equation}
  n_0 \gg \frac{m^2v^2T}{4\pi\alpha} \sim 0.1 n_{\textrm{B}}
\end{equation}
$n_{\textrm{B}} = (mT/2\pi)^{3/2}$ is the typical density in the Boltzmann
regime.  The typical electric fields in this regime are $E\sim \nabla
\phi \sim e T/2z_{\alpha}$, which yield an ionization potential of
$V_{\text{ionize}}\sim e E a_{0} \sim \alpha T a_{0}/2z_{\alpha} \sim
1$~meV~$\ll 13$~eV.  Thus, the ionization of incoming hydrogen will
not occur until the density is much larger, by which point the
screening will be highly efficient.  Contrary to the arguments
presented in~\cite{Cumberbatch:2006bj}, electrons depositive via
neutral hydrogen can easily penetrate deeply into the electrosphere
producing the direct-annihilation emissions discussed in
Sec.~\ref{sec:diffuse-1-20}.

Finally, we point out that, in the ultrarelativistic regime, we may
express~(\ref{eq:ratio-1}) in terms of the Fermi momentum:
\begin{equation}
  \label{eq:ratio-2}
  \frac{\lambda}{\lambda_D}\sim\frac{p_F}{p}\sqrt{\frac{
      \alpha}{\pi}}.
\end{equation}
Thus, for highly energetic particles, screening is irrelevant.  The
would apply to the $\sim 5$~MeV positrons ejected from the core of the
nuggets that we argued are responsible for the diffuse 10~keV
emission~\cite{Forbes:2006ba}.  These see the full electric
field which is responsible for their deceleration and ultimately for
the emission of their energy in the 10~keV band.  This point is somewhat
irrelevant, however, as the positron electrosphere of the antimatter
nuggets cannot screen a positive charge (screening is only efficient
for particles of the opposite charge than the ion constituents).


%
\end{fmffile}


\begin{thebibliography}{10}%
\makeatletter
\providecommand \@ifxundefined [1]{%
 \ifx #1\undefined \expandafter \@firstoftwo
 \else \expandafter \@secondoftwo
\fi
}%
\providecommand \@ifnum [1]{%
 \ifnum #1\expandafter \@firstoftwo
 \else \expandafter \@secondoftwo
\fi
}%
\providecommand \enquote [1]{``#1''}%
\providecommand \bibnamefont  [1]{#1}%
\providecommand \bibfnamefont [1]{#1}%
\providecommand \citenamefont [1]{#1}%
\providecommand\href[0]{\@sanitize\@href}%
\providecommand\@href[1]{\endgroup\@@startlink{#1}\endgroup\@@href}%
\providecommand\@@href[1]{#1\@@endlink}%
\providecommand \@sanitize [0]{\begingroup\catcode`\&12\catcode`\#12\relax}%
\@ifxundefined \pdfoutput {\@firstoftwo}{%
 \@ifnum{\z@=\pdfoutput}{\@firstoftwo}{\@secondoftwo}%
}{%
 \providecommand\@@startlink[1]{\leavevmode}%
 \providecommand\@@endlink[0]{}%
}{%
 \providecommand\@@startlink[1]{%
  \leavevmode
  \pdfstartlink
   attr{/Border[0 0 1 ]/H/I/C[0 1 1]}%
   user{/Subtype/Link/A<</Type/Action/S/URI/URI(#1)>>}%
  \relax
 }%
 \providecommand\@@endlink[0]{\pdfendlink}%
}%
\providecommand \url  [0]{\begingroup\@sanitize \@url }%
\providecommand \@url [1]{\endgroup\@href {#1}{\urlprefix}}%
\providecommand \urlprefix [0]{URL }%
\providecommand \Eprint[0]{\href }%
\@ifxundefined \urlstyle {%
  \providecommand \doi [1]{doi:\discretionary{}{}{}#1}%
}{%
  \providecommand \doi [0]{doi:\discretionary{}{}{}\begingroup
  \urlstyle{rm}\Url }%
}%
\providecommand \doibase [0]{http://dx.doi.org/}%
\providecommand \Doi[1]{\href{\doibase#1}}%
\providecommand \bibAnnote [3]{%
  \BibitemShut{#1}%
  \begin{quotation}\noindent
    \textsc{Key:}\ #2\\\textsc{Annotation:}\ #3%
  \end{quotation}%
}%
\providecommand \bibAnnoteFile [2]{%
  \IfFileExists{#2}{\bibAnnote {#1} {#2} {\input{#2}}}{}%
}%
\providecommand \typeout [0]{\immediate \write \m@ne }%
\providecommand \selectlanguage [0]{\@gobble}%
\providecommand \bibinfo [0]{\@secondoftwo}%
\providecommand \bibfield [0]{\@secondoftwo}%
\providecommand \translation [1]{[#1]}%
\providecommand \BibitemOpen[0]{}%
\providecommand \bibitemStop [0]{}%
\providecommand \bibitemNoStop [0]{.\EOS\space}%
\providecommand \EOS [0]{\spacefactor3000\relax}%
\providecommand \BibitemShut [1]{\csname bibitem#1\endcsname}%
\bibitem{Zhitnitsky:2002qa}%
  \BibitemOpen
  \bibfield{author}{%
  \bibinfo {author} {\bibfnamefont{A.~R.}\ \bibnamefont{Zhitnitsky}},\ }%
  \bibfield{journal}{%
  \bibinfo {journal} {\textsc{jcap}}\ }%
  \textbf{\bibinfo {volume} {10}},\ \bibinfo {pages} {010} (\bibinfo {month}
  {May}\ \bibinfo {year} {2003}),\
  \Eprint{http://arxiv.org/abs/arXiv:hep-ph/0202161}{arXiv:hep-ph/0202161}%
  \bibAnnoteFile{NoStop}{Zhitnitsky:2002qa}%
\bibitem{Oaknin:2003uv}%
  \BibitemOpen
  \bibfield{author}{%
  \bibinfo {author} {\bibfnamefont{D.~H.}\ \bibnamefont{Oaknin}}\ and\ \bibinfo
  {author} {\bibfnamefont{A.}~\bibnamefont{Zhitnitsky}},\ }%
  \bibfield{journal}{%
  \bibinfo {journal} {Phys. Rev. D}\ }%
  \textbf{\bibinfo {volume} {71}},\ \bibinfo {pages} {023519} (\bibinfo {month}
  {Jan.}\ \bibinfo {year} {2005}),\
  \Eprint{http://arxiv.org/abs/arXiv:hep-ph/0309086}{arXiv:hep-ph/0309086}%
  \bibAnnoteFile{NoStop}{Oaknin:2003uv}%
\bibitem{Zhitnitsky:2006vt}%
  \BibitemOpen
  \bibfield{author}{%
  \bibinfo {author} {\bibfnamefont{A.}~\bibnamefont{Zhitnitsky}},\ }%
  \bibfield{journal}{%
  \bibinfo {journal} {Phys. Rev. D}\ }%
  \textbf{\bibinfo {volume} {74}},\ \bibinfo {pages} {043515} (\bibinfo {month}
  {Aug.}\ \bibinfo {year} {2006}),\
  \Eprint{http://arxiv.org/abs/arXiv:astro-ph/0603064}{arXiv:astro-ph/0603064}%
  \bibAnnoteFile{NoStop}{Zhitnitsky:2006vt}%
\bibitem{Forbes:2006ba}%
  \BibitemOpen
  \bibfield{author}{%
  \bibinfo {author} {\bibfnamefont{M.~M.}\ \bibnamefont{Forbes}}\ and\ \bibinfo
  {author} {\bibfnamefont{A.~R.}\ \bibnamefont{Zhitnitsky}},\ }%
  \bibfield{journal}{%
  \bibinfo {journal} {\textsc{jcap}}\ }%
  \textbf{\bibinfo {volume} {0801}},\ \bibinfo {pages} {023} (\bibinfo {year}
  {2008}),\
  \Eprint{http://arxiv.org/abs/arXiv:astro-ph/0611506}{arXiv:astro-ph/0611506}%
  \bibAnnoteFile{NoStop}{Forbes:2006ba}%
\bibitem{Forbes:2008uf}%
  \BibitemOpen
  \bibfield{author}{%
  \bibinfo {author} {\bibfnamefont{M.~M.}\ \bibnamefont{Forbes}}\ and\ \bibinfo
  {author} {\bibfnamefont{A.~R.}\ \bibnamefont{Zhitnitsky}},\ }%
  \bibfield{journal}{%
  \bibinfo {journal} {Phys. Rev. D}\ }%
  \textbf{\bibinfo {volume} {78}},\ \bibinfo {pages} {083505} (\bibinfo {year}
  {2008}),\ \Eprint{http://arxiv.org/abs/0802.3830}{arXiv:0802.3830
  [astro-ph]}%
  \bibAnnoteFile{NoStop}{Forbes:2008uf}%
\bibitem{Lawson:2007kp}%
  \BibitemOpen
  \bibfield{author}{%
  \bibinfo {author} {\bibfnamefont{K.}~\bibnamefont{Lawson}}\ and\ \bibinfo
  {author} {\bibfnamefont{A.~R.}\ \bibnamefont{Zhitnitsky}},\ }%
  \bibfield{journal}{%
  \bibinfo {journal} {\textsc{jcap}}\ }%
  \textbf{\bibinfo {volume} {0801}},\ \bibinfo {pages} {022} (\bibinfo {month}
  {January}\ \bibinfo {year} {2008}),\
  \Eprint{http://arxiv.org/abs/arXiv:0704.3064 [astro-ph]}{arXiv:0704.3064
  [astro-ph]}%
  \bibAnnoteFile{NoStop}{Lawson:2007kp}%
\bibitem{Witten:1984rs}%
  \BibitemOpen
  \bibfield{author}{%
  \bibinfo {author} {\bibfnamefont{E.}~\bibnamefont{Witten}},\ }%
  \bibfield{journal}{%
  \bibinfo {journal} {Phys. Rev. D}\ }%
  \textbf{\bibinfo {volume} {30}},\ \bibinfo {pages} {272} (\bibinfo {year}
  {1984})%
  \bibAnnoteFile{NoStop}{Witten:1984rs}%
\bibitem{Peccei:1977ur}%
  \BibitemOpen
  \bibfield{author}{%
  \bibinfo {author} {\bibfnamefont{R.~D.}\ \bibnamefont{Peccei}}\ and\ \bibinfo
  {author} {\bibfnamefont{H.~R.}\ \bibnamefont{Quinn}},\ }%
  \bibfield{journal}{%
  \bibinfo {journal} {Phys. Rev. D}\ }%
  \textbf{\bibinfo {volume} {16}},\ \bibinfo {pages} {1791} (\bibinfo {year}
  {1977})%
  \bibAnnoteFile{NoStop}{Peccei:1977ur}%
\bibitem{Weinberg:1978ma}%
  \BibitemOpen
  \bibfield{author}{%
  \bibinfo {author} {\bibfnamefont{S.}~\bibnamefont{Weinberg}},\ }%
  \bibfield{journal}{%
  \bibinfo {journal} {Phys. Rev. Lett.}\ }%
  \textbf{\bibinfo {volume} {40}},\ \bibinfo {pages} {223} (\bibinfo {year}
  {1978})%
  \bibAnnoteFile{NoStop}{Weinberg:1978ma}%
\bibitem{Wilczek:1978pj}%
  \BibitemOpen
  \bibfield{author}{%
  \bibinfo {author} {\bibfnamefont{F.}~\bibnamefont{Wilczek}},\ }%
  \bibfield{journal}{%
  \bibinfo {journal} {Phys. Rev. Lett.}\ }%
  \textbf{\bibinfo {volume} {40}},\ \bibinfo {pages} {279} (\bibinfo {year}
  {1978})%
  \bibAnnoteFile{NoStop}{Wilczek:1978pj}%
\bibitem{Kim:1979if}%
  \BibitemOpen
  \bibfield{author}{%
  \bibinfo {author} {\bibfnamefont{J.~E.}\ \bibnamefont{Kim}},\ }%
  \bibfield{journal}{%
  \bibinfo {journal} {Phys. Rev. Lett.}\ }%
  \textbf{\bibinfo {volume} {43}},\ \bibinfo {pages} {103} (\bibinfo {year}
  {1979})%
  \bibAnnoteFile{NoStop}{Kim:1979if}%
\bibitem{Shifman:1980if}%
  \BibitemOpen
  \bibfield{author}{%
  \bibinfo {author} {\bibfnamefont{M.~A.}\ \bibnamefont{Shifman}}, \bibinfo
  {author} {\bibfnamefont{A.~I.}\ \bibnamefont{Vainshtein}},\ and\ \bibinfo
  {author} {\bibfnamefont{V.~I.}\ \bibnamefont{Zakharov}},\ }%
  \bibfield{journal}{%
  \bibinfo {journal} {Nucl. Phys.}\ }%
  \textbf{\bibinfo {volume} {B166}},\ \bibinfo {pages} {493} (\bibinfo {year}
  {1980})%
  \bibAnnoteFile{NoStop}{Shifman:1980if}%
\bibitem{Dine:1981rt}%
  \BibitemOpen
  \bibfield{author}{%
  \bibinfo {author} {\bibfnamefont{M.}~\bibnamefont{Dine}}, \bibinfo {author}
  {\bibfnamefont{W.}~\bibnamefont{Fischler}},\ and\ \bibinfo {author}
  {\bibfnamefont{M.}~\bibnamefont{Srednicki}},\ }%
  \bibfield{journal}{%
  \bibinfo {journal} {Phys. Lett.}\ }%
  \textbf{\bibinfo {volume} {B104}},\ \bibinfo {pages} {199} (\bibinfo {year}
  {1981})%
  \bibAnnoteFile{NoStop}{Dine:1981rt}%
\bibitem{Zhitnitsky:1980tq}%
  \BibitemOpen
  \bibfield{author}{%
  \bibinfo {author} {\bibfnamefont{A.~R.}\ \bibnamefont{Zhitnitsky}},\ }%
  \bibfield{journal}{%
  \bibinfo {journal} {Sov. J. Nucl. Phys.}\ }%
  \textbf{\bibinfo {volume} {31}},\ \bibinfo {pages} {260} (\bibinfo {year}
  {1980})%
  \bibAnnoteFile{NoStop}{Zhitnitsky:1980tq}%
\bibitem{Srednicki:2002ww}%
  \BibitemOpen
  \bibfield{author}{%
  \bibinfo {author} {\bibfnamefont{M.}~\bibnamefont{Srednicki}}}
   (\bibinfo {year} {2002}),\
  \Eprint{http://arxiv.org/abs/arXiv:hep-th/0210172}{arXiv:hep-th/0210172}%
  \bibAnnoteFile{NoStop}{Srednicki:2002ww}%
\bibitem{vanBibber:2006rb}%
  \BibitemOpen
  \bibfield{author}{%
  \bibinfo {author} {\bibfnamefont{K.}~\bibnamefont{van Bibber}}\ and\ \bibinfo
  {author} {\bibfnamefont{L.~J.}\ \bibnamefont{Rosenberg}},\ }%
  \bibfield{journal}{%
  \bibinfo {journal} {Phys. Today}\ }%
  \textbf{\bibinfo {volume} {59N8}},\ \bibinfo {pages} {30} (\bibinfo {year}
  {2006})%
  \bibAnnoteFile{NoStop}{vanBibber:2006rb}%
\bibitem{Asztalos:2006kz}%
  \BibitemOpen
  \bibfield{author}{%
  \bibinfo {author} {\bibfnamefont{S.~J.}\ \bibnamefont{Asztalos}}, \bibinfo
  {author} {\bibfnamefont{L.~J.}\ \bibnamefont{Rosenberg}}, \bibinfo {author}
  {\bibfnamefont{K.}~\bibnamefont{van Bibber}}, \bibinfo {author}
  {\bibfnamefont{P.}~\bibnamefont{Sikivie}},\ and\ \bibinfo {author}
  {\bibfnamefont{K.}~\bibnamefont{Zioutas}},\ }%
  \bibfield{journal}{%
  \bibinfo {journal} {Ann. Rev. Nucl. Part. Sci.}\ }%
  \textbf{\bibinfo {volume} {56}},\ \bibinfo {pages} {293} (\bibinfo {year}
  {2006})%
  \bibAnnoteFile{NoStop}{Asztalos:2006kz}%
\bibitem{Dolgov:2009ix}%
  \BibitemOpen
  \bibfield{author}{%
  \bibinfo {author} {\bibfnamefont{A.}~\bibnamefont{Dolgov}}\ and\ \bibinfo
  {author} {\bibfnamefont{A.~R.}\ \bibnamefont{Zhitnitsky}},\ }%
  \bibinfo {howpublished} {work in progress} (\bibinfo {year} {2009})%
  \bibAnnoteFile{NoStop}{Dolgov:2009ix}%
\bibitem{Kharzeev:2007tn}%
  \BibitemOpen
  \bibfield{author}{%
  \bibinfo {author} {\bibfnamefont{D.}~\bibnamefont{Kharzeev}}\ and\ \bibinfo
  {author} {\bibfnamefont{A.}~\bibnamefont{Zhitnitsky}},\ }%
  \bibfield{journal}{%
  \bibinfo {journal} {Nucl. Phys.}\ }%
  \textbf{\bibinfo {volume} {A797}},\ \bibinfo {pages} {67} (\bibinfo {year}
  {2007}),\ \Eprint{http://arxiv.org/abs/arXiv:0706.1026
  [hep-ph]}{arXiv:0706.1026 [hep-ph]}%
  \bibAnnoteFile{NoStop}{Kharzeev:2007tn}%
\bibitem{Voloshin:2010ut}%
  \BibitemOpen
  \bibfield{author}{%
  \bibinfo {author} {\bibfnamefont{S.~A.}\ \bibnamefont{Voloshin}},\ }%
  \enquote{\bibinfo {title} {{T}esting chiral magnetic effect with central
  {U}+{U} collisions},}\  (\bibinfo {year} {2010}),\
  \Eprint{http://arxiv.org/abs/arXiv:1006.1020}{arXiv:1006.1020}%
  \bibAnnoteFile{NoStop}{Voloshin:2010ut}%
\bibitem{Abelev:2010tx}%
  \BibitemOpen
  \bibfield{author}{%
  \bibinfo {author} {\bibfnamefont{B.~I.}\ \bibnamefont{Abelev}},
  \textit{et al.} (\bibinfo {collaboration} {STAR
    Collaboration}),\ }%
  \bibfield{journal}{%
  \Doi{10.1103/PhysRevC.81.054908}{\bibinfo {journal} {Phys. Rev. C}}\ }%
  \textbf{\bibinfo {volume} {81}},\ \bibinfo {pages} {054908} (\bibinfo {month}
  {May}\ \bibinfo {year} {2010}),\
  \Eprint{http://arxiv.org/abs/arXiv:0909.1717}{arXiv:0909.1717}%
  \bibAnnoteFile{NoStop}{Abelev:2010tx}%
\bibitem{Amsler:2008zz}%
  \BibitemOpen
  \bibfield{author}{%
  \bibinfo {author} {\bibfnamefont{C.}~\bibnamefont{Amsler}} \emph{et~al.}
  (\bibinfo {collaboration} {Particle Data Group}),\ }%
  \bibfield{journal}{%
  \bibinfo {journal} {Phys. Lett.}\ }%
  \textbf{\bibinfo {volume} {B667}},\ \bibinfo {pages} {1} (\bibinfo {year}
  {2008})%
  \bibAnnoteFile{NoStop}{Amsler:2008zz}%
\bibitem{Herrin:2005kb}%
  \BibitemOpen
  \bibfield{author}{%
  \bibinfo {author} {\bibfnamefont{E.~T.}\ \bibnamefont{Herrin}}, \bibinfo
  {author} {\bibfnamefont{D.~C.}\ \bibnamefont{Rosenbaum}},\ and\ \bibinfo
  {author} {\bibfnamefont{V.~L.}\ \bibnamefont{Teplitz}},\ }%
  \bibfield{journal}{%
  \bibinfo {journal} {Phys. Rev. D}\ }%
  \textbf{\bibinfo {volume} {73}},\ \bibinfo {pages} {043511} (\bibinfo {year}
  {2006}),\
  \Eprint{http://arxiv.org/abs/arXiv:astro-ph/0505584}{arXiv:astro-ph/0505584}%
  \bibAnnoteFile{NoStop}{Herrin:2005kb}%
\bibitem{Abers:2007ji}%
  \BibitemOpen
  \bibfield{author}{%
  \bibinfo {author} {\bibfnamefont{E.~S.}\ \bibnamefont{Abers}}, \bibinfo
  {author} {\bibfnamefont{A.~K.}\ \bibnamefont{Bhatia}}, \bibinfo {author}
  {\bibfnamefont{D.~A.}\ \bibnamefont{Dicus}}, \bibinfo {author}
  {\bibfnamefont{W.~W.}\ \bibnamefont{Repko}}, \bibinfo {author}
  {\bibfnamefont{D.~C.}\ \bibnamefont{Rosenbaum}},\ and\ \bibinfo {author}
  {\bibfnamefont{V.~L.}\ \bibnamefont{Teplitz}},\ }%
  \bibfield{journal}{%
  \bibinfo {journal} {Phys. Rev. D}\ }%
  \textbf{\bibinfo {volume} {79}},\ \bibinfo {pages} {023513} (\bibinfo {year}
  {2009}),\ \Eprint{http://arxiv.org/abs/arXiv:0712.4300
  [astro-ph]}{arXiv:0712.4300 [astro-ph]}%
  \bibAnnoteFile{NoStop}{Abers:2007ji}%
\bibitem{kolb94:_early_univer}%
  \BibitemOpen
  \bibfield{author}{%
  \bibinfo {author} {\bibfnamefont{E.~W.}\ \bibnamefont{Kolb}}\ and\ \bibinfo
  {author} {\bibfnamefont{M.~S.}\ \bibnamefont{Turner}},\ }%
  \emph{\bibinfo {title} {The Early Universe}}\ (\bibinfo {publisher} {Westview
  Press},\ \bibinfo {address} {5500 Central Avenue, Boulder, Colorado, 80301},\
  \bibinfo {year} {1994})%
  \bibAnnoteFile{NoStop}{kolb94:_early_univer}%
\bibitem{Knodlseder:2003sv}%
  \BibitemOpen
  \bibfield{author}{%
  \bibinfo {author} {\bibfnamefont{J.}~\bibnamefont{{Kn\"odlseder}}}, \bibinfo
  {author} {\bibfnamefont{V.}~\bibnamefont{Lonjou}}, \bibinfo {author}
  {\bibfnamefont{P.}~\bibnamefont{Jean}}, \bibinfo {author}
  {\bibfnamefont{M.}~\bibnamefont{Allain}}, \bibinfo {author}
  {\bibfnamefont{P.}~\bibnamefont{Mandrou}}, \bibinfo {author}
  {\bibfnamefont{J.-P.}\ \bibnamefont{Roques}}, \bibinfo {author}
  {\bibfnamefont{G.}~\bibnamefont{Skinner}}, \bibinfo {author}
  {\bibfnamefont{G.}~\bibnamefont{Vedrenne}}, \bibinfo {author}
  {\bibfnamefont{P.}~\bibnamefont{von Ballmoos}}, \bibinfo {author}
  {\bibfnamefont{G.}~\bibnamefont{Weidenspointner}}, \bibinfo {author}
  {\bibfnamefont{P.}~\bibnamefont{Caraveo}}, \bibinfo {author}
  {\bibfnamefont{B.}~\bibnamefont{Cordier}}, \bibinfo {author}
  {\bibfnamefont{V.}~\bibnamefont{{Sch\"onfelder}}},\ and\ \bibinfo {author}
  {\bibfnamefont{B.}~\bibnamefont{Teegarden}},\ }%
  \bibfield{journal}{%
  \bibinfo {journal} {Astron. Astrophys.}\ }%
  \textbf{\bibinfo {volume} {411}},\ \bibinfo {pages} {L457} (\bibinfo {year}
  {2003}),\
  \Eprint{http://arxiv.org/abs/arXiv:astro-ph/0309442}{arXiv:astro-ph/0309442}%
  \bibAnnoteFile{NoStop}{Knodlseder:2003sv}%
\bibitem{Beacom:2005qv}%
  \BibitemOpen
  \bibfield{author}{%
  \bibinfo {author} {\bibfnamefont{J.~F.}\ \bibnamefont{Beacom}}\ and\ \bibinfo
  {author} {\bibfnamefont{H.}~\bibnamefont{{Y\"uksel}}},\ }%
  \bibfield{journal}{%
  \bibinfo {journal} {Phys. Rev. Lett.}\ }%
  \textbf{\bibinfo {volume} {97}},\ \bibinfo {pages} {071102} (\bibinfo {year}
  {2006}),\
  \Eprint{http://arxiv.org/abs/arXiv:astro-ph/0512411}{arXiv:astro-ph/0512411}%
  \bibAnnoteFile{NoStop}{Beacom:2005qv}%
\bibitem{Yuksel:2006fj}%
  \BibitemOpen
  \bibfield{author}{%
  \bibinfo {author} {\bibfnamefont{H.}~\bibnamefont{{Y\"uksel}}},\ }%
  \bibfield{journal}{%
  \bibinfo {journal} {Nucl. Phys. B Proc. Suppl.}\ }%
  \textbf{\bibinfo {volume} {173}},\ \bibinfo {pages} {83} (\bibinfo {year}
  {2006}),\
  \Eprint{http://arxiv.org/abs/arXiv:astro-ph/0609139}{arXiv:astro-ph/0609139}%
  \bibAnnoteFile{NoStop}{Yuksel:2006fj}%
\bibitem{Oaknin:2004mn}%
  \BibitemOpen
  \bibfield{author}{%
  \bibinfo {author} {\bibfnamefont{D.~H.}\ \bibnamefont{Oaknin}}\ and\ \bibinfo
  {author} {\bibfnamefont{A.~R.}\ \bibnamefont{Zhitnitsky}},\ }%
  \bibfield{journal}{%
  \bibinfo {journal} {Phys. Rev. Lett.}\ }%
  \textbf{\bibinfo {volume} {94}},\ \bibinfo {pages} {101301} (\bibinfo {year}
  {2005}),\
  \Eprint{http://arxiv.org/abs/arXiv:hep-ph/0406146}{arXiv:hep-ph/0406146}%
  \bibAnnoteFile{NoStop}{Oaknin:2004mn}%
\bibitem{Zhitnitsky:2006tu}%
  \BibitemOpen
  \bibfield{author}{%
  \bibinfo {author} {\bibfnamefont{A.}~\bibnamefont{Zhitnitsky}},\ }%
  \bibfield{journal}{%
  \bibinfo {journal} {Phys. Rev. D}\ }%
  \textbf{\bibinfo {volume} {76}},\ \bibinfo {pages} {103518} (\bibinfo {year}
  {2007}),\
  \Eprint{http://arxiv.org/abs/arXiv:astro-ph/0607361}{arXiv:astro-ph/0607361}%
  \bibAnnoteFile{NoStop}{Zhitnitsky:2006tu}%
\bibitem{Muno:2004bs}%
  \BibitemOpen
  \bibfield{author}{%
  \bibinfo {author} {\bibfnamefont{M.~P.}\ \bibnamefont{Muno}} \emph{et~al.},\
  }%
  \bibfield{journal}{%
  \bibinfo {journal} {Astrophys. J.}\ }%
  \textbf{\bibinfo {volume} {613}},\ \bibinfo {pages} {326} (\bibinfo {year}
  {2004}),\
  \Eprint{http://arxiv.org/abs/arXiv:astro-ph/0402087}{arXiv:astro-ph/0402087}%
  \bibAnnoteFile{NoStop}{Muno:2004bs}%
\bibitem{Finkbeiner:2003im}%
  \BibitemOpen
  \bibfield{author}{%
  \bibinfo {author} {\bibfnamefont{D.~P.}\ \bibnamefont{Finkbeiner}},\ }%
  \bibfield{journal}{%
  \bibinfo {journal} {Astrophys. J.}\ }%
  \textbf{\bibinfo {volume} {614}},\ \bibinfo {pages} {186} (\bibinfo {year}
  {2004}),\
  \Eprint{http://arxiv.org/abs/arXiv:astro-ph/0311547}{arXiv:astro-ph/0311547}%
  \bibAnnoteFile{NoStop}{Finkbeiner:2003im}%
\bibitem{Finkbeiner:2004je}%
  \BibitemOpen
  \bibfield{author}{%
  \bibinfo {author} {\bibfnamefont{D.~P.}\ \bibnamefont{Finkbeiner}}, \bibinfo
  {author} {\bibfnamefont{G.~I.}\ \bibnamefont{Langston}},\ and\ \bibinfo
  {author} {\bibfnamefont{A.~H.}\ \bibnamefont{Minter}},\ }%
  \bibfield{journal}{%
  \bibinfo {journal} {Astrophys. J.}\ }%
  \textbf{\bibinfo {volume} {617}},\ \bibinfo {pages} {350} (\bibinfo {year}
  {2004}),\
  \Eprint{http://arxiv.org/abs/arXiv:astro-ph/0408292}{arXiv:astro-ph/0408292}%
  \bibAnnoteFile{NoStop}{Finkbeiner:2004je}%
\bibitem{Hooper:2007gi}%
  \BibitemOpen
  \bibfield{author}{%
  \bibinfo {author} {\bibfnamefont{D.}~\bibnamefont{Hooper}}, \bibinfo {author}
  {\bibfnamefont{G.}~\bibnamefont{Zaharijas}}, \bibinfo {author}
  {\bibfnamefont{D.~P.}\ \bibnamefont{Finkbeiner}},\ and\ \bibinfo {author}
  {\bibfnamefont{G.}~\bibnamefont{Dobler}},\ }%
  \bibfield{journal}{%
  \bibinfo {journal} {Phys. Rev. D}\ }%
  \textbf{\bibinfo {volume} {77}},\ \bibinfo {pages} {043511} (\bibinfo {month}
  {February}\ \bibinfo {year} {2007}),\
  \Eprint{http://arxiv.org/abs/arXiv:0709.3114 [astro-ph]}{arXiv:0709.3114
  [astro-ph]}%
  \bibAnnoteFile{NoStop}{Hooper:2007gi}%
\bibitem{Dobler:2007wv}%
  \BibitemOpen
  \bibfield{author}{%
  \bibinfo {author} {\bibfnamefont{G.}~\bibnamefont{Dobler}}\ and\ \bibinfo
  {author} {\bibfnamefont{D.~P.}\ \bibnamefont{Finkbeiner}},\ }%
  \bibfield{journal}{%
  \bibinfo {journal} {Astrophys. J.}\ }%
  \textbf{\bibinfo {volume} {680}},\ \bibinfo {pages} {1222 } (\bibinfo {month}
  {June}\ \bibinfo {year} {2008}),\
  \Eprint{http://arxiv.org/abs/arXiv:0712.1038 [astro-ph]}{arXiv:0712.1038
  [astro-ph]}%
  \bibAnnoteFile{NoStop}{Dobler:2007wv}%
\bibitem{Lingenfelter:2009fk}%
  \BibitemOpen
  \bibfield{author}{%
  \bibinfo {author} {\bibfnamefont{R.~E.}\ \bibnamefont{Lingenfelter}},
  \bibinfo {author} {\bibfnamefont{J.~C.}\ \bibnamefont{Higdon}},\ and\
  \bibinfo {author} {\bibfnamefont{R.~E.}\ \bibnamefont{Rothschild}},\ }%
  \bibfield{journal}{%
  \bibinfo {journal} {Phys. Rev. Lett.}\ }%
  \textbf{\bibinfo {volume} {103}},\ \bibinfo {pages} {031301} (\bibinfo
  {month} {Jul}\ \bibinfo {year} {2009}),\
  \Eprint{http://arxiv.org/abs/arXiv:0904.1025}{arXiv:0904.1025}%
  \bibAnnoteFile{NoStop}{Lingenfelter:2009fk}%
\bibitem{Farhi:1984}%
  \BibitemOpen
  \bibfield{author}{%
  \bibinfo {author} {\bibfnamefont{E.}~\bibnamefont{Farhi}}\ and\ \bibinfo
  {author} {\bibfnamefont{R.~L.}\ \bibnamefont{Jaffe}},\ }%
  \bibfield{journal}{%
  \bibinfo {journal} {Phys. Rev. D}\ }%
  \textbf{\bibinfo {volume} {30}},\ \bibinfo {pages} {2379} (\bibinfo {month}
  {Dec}\ \bibinfo {year} {1984})%
  \bibAnnoteFile{NoStop}{Farhi:1984}%
\bibitem{Madsen:2006}%
  \BibitemOpen
  \bibfield{author}{%
  \bibinfo {author} {\bibfnamefont{J.}~\bibnamefont{Madsen}},\ }%
  \enquote{\bibinfo {title} {{S}trangelets, {N}uclearites, {Q}-balls--{A}
  {B}rief {O}verview},}\ \bibinfo {howpublished} {Invited talk at Workshop on
  Exotic Physics with Neutrino Telescopes, Uppsala, Sweden, Sept. 2006}
  (\bibinfo {year} {2006}),\
  \Eprint{http://arxiv.org/abs/arXiv:astro-ph/0612740v1}{arXiv:astro-ph/061274%
0v1}%
  \bibAnnoteFile{NoStop}{Madsen:2006}%
\bibitem{Madsen:2004}%
  \BibitemOpen
  \bibfield{author}{%
  \bibinfo {author} {\bibfnamefont{J.}~\bibnamefont{Madsen}},\ }%
  \bibfield{journal}{%
  \bibinfo {journal} {Journal of Physics G: Nuclear and Particle Physics}\ }%
  \textbf{\bibinfo {volume} {31}},\ \bibinfo {pages} {S833} (\bibinfo {year}
  {2005}),\
  \Eprint{http://arxiv.org/abs/arXiv:astro-ph/0411601v1}{arXiv:astro-ph/041160%
1v1}%
  \bibAnnoteFile{NoStop}{Madsen:2004}%
\bibitem{Christiansen:1997}%
  \BibitemOpen
  \bibfield{author}{%
  \bibinfo {author} {\bibfnamefont{M.~B.}\ \bibnamefont{Christiansen}}\ and\
  \bibinfo {author} {\bibfnamefont{N.~K.}\ \bibnamefont{Glendenning}},\ }%
  \bibfield{journal}{%
  \Doi{10.1103/PhysRevC.56.2858}{\bibinfo {journal} {Phys. Rev. C}}\ }%
  \textbf{\bibinfo {volume} {56}},\ \bibinfo {pages} {2858} (\bibinfo {month}
  {Nov.}\ \bibinfo {year} {1997})%
  \bibAnnoteFile{NoStop}{Christiansen:1997}%
\bibitem{Jaikumar:2006}%
  \BibitemOpen
  \bibfield{author}{%
  \bibinfo {author} {\bibfnamefont{P.}~\bibnamefont{Jaikumar}}, \bibinfo
  {author} {\bibfnamefont{S.}~\bibnamefont{Reddy}},\ and\ \bibinfo {author}
  {\bibfnamefont{A.~W.}\ \bibnamefont{Steiner}},\ }%
  \bibfield{journal}{%
  \Doi{10.1103/PhysRevLett.96.041101}{\bibinfo {journal} {Phys. Rev. Lett.}}\
  }%
  \textbf{\bibinfo {volume} {96}},\ \bibinfo {pages} {041101} (\bibinfo {month}
  {Jan.}\ \bibinfo {year} {2006}),\
  \Eprint{http://arxiv.org/abs/arXiv:nucl-th/0507055}{arXiv:nucl-th/0507055}%
  \bibAnnoteFile{NoStop}{Jaikumar:2006}%
\bibitem{Alford:2006}%
  \BibitemOpen
  \bibfield{author}{%
  \bibinfo {author} {\bibfnamefont{M.~G.}\ \bibnamefont{Alford}}, \bibinfo
  {author} {\bibfnamefont{K.}~\bibnamefont{Rajagopal}}, \bibinfo {author}
  {\bibfnamefont{S.}~\bibnamefont{Reddy}},\ and\ \bibinfo {author}
  {\bibfnamefont{A.~W.}\ \bibnamefont{Steiner}},\ }%
  \bibfield{journal}{%
  \Doi{10.1103/PhysRevD.73.114016}{\bibinfo {journal} {Phys. Rev. D}}\ }%
  \textbf{\bibinfo {volume} {73}},\ \bibinfo {pages} {114016} (\bibinfo {month}
  {Jun.}\ \bibinfo {year} {2006}),\
  \Eprint{http://arxiv.org/abs/arXiv:hep-ph/0604134}{arXiv:hep-ph/0604134}%
  \bibAnnoteFile{NoStop}{Alford:2006}%
\bibitem{Mishustin:2004xa}%
  \BibitemOpen
  \bibfield{author}{%
  \bibinfo {author} {\bibfnamefont{I.~N.}\ \bibnamefont{Mishustin}}, \bibinfo
  {author} {\bibfnamefont{L.~M.}\ \bibnamefont{Satarov}}, \bibinfo {author}
  {\bibfnamefont{T.~J.}\ \bibnamefont{Burvenich}}, \bibinfo {author}
  {\bibfnamefont{H.}~\bibnamefont{Stoecker}},\ and\ \bibinfo {author}
  {\bibfnamefont{W.}~\bibnamefont{Greiner}},\ }%
  \bibfield{journal}{%
  \bibinfo {journal} {Phys. Rev. C}\ }%
  \textbf{\bibinfo {volume} {71}},\ \bibinfo {pages} {035201} (\bibinfo {month}
  {Mar.}\ \bibinfo {year} {2005}),\
  \Eprint{http://arxiv.org/abs/arXiv:nucl-th/0404026}{arXiv:nucl-th/0404026}%
  \bibAnnoteFile{NoStop}{Mishustin:2004xa}%
\bibitem{Larionov:2008wy}%
  \BibitemOpen
  \bibfield{author}{%
  \bibinfo {author} {\bibfnamefont{A.~B.}\ \bibnamefont{Larionov}}, \bibinfo
  {author} {\bibfnamefont{I.~N.}\ \bibnamefont{Mishustin}}, \bibinfo {author}
  {\bibfnamefont{L.~M.}\ \bibnamefont{Satarov}},\ and\ \bibinfo {author}
  {\bibfnamefont{W.}~\bibnamefont{Greiner}},\ }%
  \bibfield{journal}{%
  \Doi{10.1103/PhysRevC.78.014604}{\bibinfo {journal} {Phys. Rev. C}}\ }%
  \textbf{\bibinfo {volume} {78}},\ \bibinfo {pages} {014604} (\bibinfo {month}
  {Jul.}\ \bibinfo {year} {2008}),\
  \Eprint{http://arxiv.org/abs/arXiv:0802.1845 [nucl-th]}{arXiv:0802.1845
  [nucl-th]}%
  \bibAnnoteFile{NoStop}{Larionov:2008wy}%
\bibitem{Alcock:1986hz}%
  \BibitemOpen
  \bibfield{author}{%
  \bibinfo {author} {\bibfnamefont{C.}~\bibnamefont{Alcock}}, \bibinfo {author}
  {\bibfnamefont{E.}~\bibnamefont{Farhi}},\ and\ \bibinfo {author}
  {\bibfnamefont{A.}~\bibnamefont{Olinto}},\ }%
  \bibfield{journal}{%
  \bibinfo {journal} {Astrophys. J.}\ }%
  \textbf{\bibinfo {volume} {310}},\ \bibinfo {pages} {261} (\bibinfo {year}
  {1986})%
  \bibAnnoteFile{NoStop}{Alcock:1986hz}%
\bibitem{Kettner:1994zs}%
  \BibitemOpen
  \bibfield{author}{%
  \bibinfo {author} {\bibfnamefont{C.}~\bibnamefont{Kettner}}, \bibinfo
  {author} {\bibfnamefont{F.}~\bibnamefont{Weber}}, \bibinfo {author}
  {\bibfnamefont{M.~K.}\ \bibnamefont{Weigel}},\ and\ \bibinfo {author}
  {\bibfnamefont{N.~K.}\ \bibnamefont{Glendenning}},\ }%
  \bibfield{journal}{%
  \bibinfo {journal} {Phys. Rev. D}\ }%
  \textbf{\bibinfo {volume} {51}},\ \bibinfo {pages} {1440} (\bibinfo {year}
  {1995})%
  \bibAnnoteFile{NoStop}{Kettner:1994zs}%
\bibitem{Cheng:2003le}%
  \BibitemOpen
  \bibfield{author}{%
  \bibinfo {author} {\bibfnamefont{K.~S.}\ \bibnamefont{Cheng}}\ and\ \bibinfo
  {author} {\bibfnamefont{T.}~\bibnamefont{Harko}},\ }%
  \bibfield{journal}{%
  \bibinfo {journal} {Astrophys. J.}\ }%
  \textbf{\bibinfo {volume} {596}},\ \bibinfo {pages} {451} (\bibinfo {year}
  {2003}),\
  \Eprint{http://arxiv.org/abs/arXiv:astro-ph/0306482}{arXiv:astro-ph/0306482}%
  \bibAnnoteFile{NoStop}{Cheng:2003le}%
\bibitem{Usov:2004kj}%
  \BibitemOpen
  \bibfield{author}{%
  \bibinfo {author} {\bibfnamefont{V.~V.}\ \bibnamefont{Usov}}, \bibinfo
  {author} {\bibfnamefont{T.}~\bibnamefont{Harko}},\ and\ \bibinfo {author}
  {\bibfnamefont{K.~S.}\ \bibnamefont{Cheng}},\ }%
  \bibfield{journal}{%
  \bibinfo {journal} {Astrophys. J.}\ }%
  \textbf{\bibinfo {volume} {620}},\ \bibinfo {pages} {915} (\bibinfo {year}
  {2005}),\
  \Eprint{http://arxiv.org/abs/arXiv:astro-ph/0410682}{arXiv:astro-ph/0410682}%
  \bibAnnoteFile{NoStop}{Usov:2004kj}%
\bibitem{Heiselberg:1993}%
  \BibitemOpen
  \bibfield{author}{%
  \bibinfo {author} {\bibfnamefont{H.}~\bibnamefont{Heiselberg}},\ }%
  \bibfield{journal}{%
  \Doi{10.1103/PhysRevD.48.1418}{\bibinfo {journal} {Phys. Rev. D}}\ }%
  \textbf{\bibinfo {volume} {48}},\ \bibinfo {pages} {1418} (\bibinfo {month}
  {Aug.}\ \bibinfo {year} {1993})%
  \bibAnnoteFile{NoStop}{Heiselberg:1993}%
\bibitem{LL3:1977}%
  \BibitemOpen
  \bibfield{author}{%
  \bibinfo {author} {\bibfnamefont{L.~D.}\ \bibnamefont{Landau}}\ and\ \bibinfo
  {author} {\bibfnamefont{E.~M.}\ \bibnamefont{Lifshitz}},\ }%
  \emph{\bibinfo {title} {Quantum Mechanics: Non-relativistic theory}},\
  \bibinfo {edition} {3rd}\ ed.,\ \bibinfo {series} {Course of Theoretical
  Physics}, Vol.~\bibinfo {volume} {3}\ (\bibinfo {publisher}
  {Butterworth-Heinemann},\ \bibinfo {address} {Oxford},\ \bibinfo {year}
  {2003, c1977})%
  \bibAnnoteFile{NoStop}{LL3:1977}%
\bibitem{Cumberbatch:2006bj}%
  \BibitemOpen
  \bibfield{author}{%
  \bibinfo {author} {\bibfnamefont{D.~T.}\ \bibnamefont{Cumberbatch}}, \bibinfo
  {author} {\bibfnamefont{J.}~\bibnamefont{Silk}},\ and\ \bibinfo {author}
  {\bibfnamefont{G.~D.}\ \bibnamefont{Starkman}},\ }%
  \bibfield{journal}{%
  \bibinfo {journal} {Phys. Rev. D}\ }%
  \textbf{\bibinfo {volume} {77}},\ \bibinfo {pages} {063522} (\bibinfo {month}
  {March}\ \bibinfo {year} {2008}),\
  \Eprint{http://arxiv.org/abs/arXiv:astro-ph/0606429}{arXiv:astro-ph/0606429}%
  \bibAnnoteFile{NoStop}{Cumberbatch:2006bj}%
\bibitem{Usov:2004iz}%
  \BibitemOpen
  \bibfield{author}{%
  \bibinfo {author} {\bibfnamefont{V.~V.}\ \bibnamefont{Usov}},\ }%
  \bibfield{journal}{%
  \bibinfo {journal} {Phys. Rev. D}\ }%
  \textbf{\bibinfo {volume} {70}},\ \bibinfo {pages} {067301} (\bibinfo {month}
  {September}\ \bibinfo {year} {2004}),\
  \Eprint{http://arxiv.org/abs/arXiv:astro-ph/0408217}{arXiv:astro-ph/0408217}%
  \bibAnnoteFile{NoStop}{Usov:2004iz}%
\bibitem{Ferriere:2007rz}%
  \BibitemOpen
  \bibfield{author}{%
  \bibinfo {author} {\bibfnamefont{K.}~\bibnamefont{{Ferri\`ere}}}, \bibinfo
  {author} {\bibfnamefont{W.}~\bibnamefont{Gillard}},\ and\ \bibinfo {author}
  {\bibfnamefont{P.}~\bibnamefont{Jean}},\ }%
  \bibfield{journal}{%
  \bibinfo {journal} {Astron. \& Astrophys.}\ }%
  \textbf{\bibinfo {volume} {467}},\ \bibinfo {pages} {611} (\bibinfo {year}
  {2007}),\
  \Eprint{http://arxiv.org/abs/arXiv:astro-ph/0702532}{arXiv:astro-ph/0702532}%
  \bibAnnoteFile{NoStop}{Ferriere:2007rz}%
\bibitem{Madsen:2008}%
  \BibitemOpen
  \bibfield{author}{%
  \bibinfo {author} {\bibfnamefont{J.}~\bibnamefont{Madsen}},\ }%
  \bibfield{journal}{%
  \bibinfo {journal} {Phys. Rev. Lett.}\ }%
  \textbf{\bibinfo {volume} {100}},\ \bibinfo {pages} {151102} (\bibinfo
  {month} {Apr.}\ \bibinfo {year} {2008}),\
  \Eprint{http://arxiv.org/abs/arXiv:0804.2140}{arXiv:0804.2140}%
  \bibAnnoteFile{NoStop}{Madsen:2008}%
\bibitem{Weidenspointner:2006nu}%
  \BibitemOpen
  \bibfield{author}{%
  \bibinfo {author} {\bibfnamefont{G.}~\bibnamefont{Weidenspointner}}, \bibinfo
  {author} {\bibfnamefont{C.~R.}\ \bibnamefont{Shrader}}, \bibinfo {author}
  {\bibfnamefont{J.}~\bibnamefont{{Kn\"odlseder}}}, \bibinfo {author}
  {\bibfnamefont{P.}~\bibnamefont{Jean}}, \bibinfo {author}
  {\bibfnamefont{V.}~\bibnamefont{Lonjou}}, \bibinfo {author}
  {\bibfnamefont{N.}~\bibnamefont{Guessoum}}, \bibinfo {author}
  {\bibfnamefont{R.}~\bibnamefont{Diehl}}, \bibinfo {author}
  {\bibfnamefont{W.}~\bibnamefont{Gillard}}, \bibinfo {author}
  {\bibfnamefont{M.}~\bibnamefont{Harris}}, \bibinfo {author}
  {\bibfnamefont{G.}~\bibnamefont{Skinner}}, \bibinfo {author}
  {\bibfnamefont{P.}~\bibnamefont{von Ballmoos}}, \bibinfo {author}
  {\bibfnamefont{G.}~\bibnamefont{Vedrenne}}, \bibinfo {author}
  {\bibfnamefont{J.-P.}\ \bibnamefont{Roques}}, \bibinfo {author}
  {\bibfnamefont{S.}~\bibnamefont{Schanne}}, \bibinfo {author}
  {\bibfnamefont{P.}~\bibnamefont{Sizun}}, \bibinfo {author}
  {\bibfnamefont{B.}~\bibnamefont{Teegarden}}, \bibinfo {author}
  {\bibfnamefont{V.}~\bibnamefont{Schoenfelder}},\ and\ \bibinfo {author}
  {\bibfnamefont{C.}~\bibnamefont{Winkler}},\ }%
  \bibfield{journal}{%
  \bibinfo {journal} {Astron. Astrophys.}\ }%
  \textbf{\bibinfo {volume} {450}},\ \bibinfo {pages} {1013} (\bibinfo {year}
  {2006}),\
  \Eprint{http://arxiv.org/abs/arXiv:astro-ph/0601673}{arXiv:astro-ph/0601673}%
  \bibAnnoteFile{NoStop}{Weidenspointner:2006nu}%
\bibitem{2008Natur.451..159W}%
  \BibitemOpen
  \bibfield{author}{%
  \bibinfo {author} {\bibfnamefont{G.}~\bibnamefont{Weidenspointner}}, \bibinfo
  {author} {\bibfnamefont{G.~K.}\ \bibnamefont{Skinner}}, \bibinfo {author}
  {\bibfnamefont{P.}~\bibnamefont{Jean}}, \bibinfo {author}
  {\bibfnamefont{J.}~\bibnamefont{{Kn\"odlseder}}}, \bibinfo {author}
  {\bibfnamefont{P.}~\bibnamefont{von Ballmoos}}, \bibinfo {author}
  {\bibfnamefont{G.}~\bibnamefont{Bignam}}, \bibinfo {author}
  {\bibfnamefont{R.}~\bibnamefont{Diehl}}, \bibinfo {author}
  {\bibfnamefont{A.~W.}\ \bibnamefont{Strong}}, \bibinfo {author}
  {\bibfnamefont{B.}~\bibnamefont{Cordier}}, \bibinfo {author}
  {\bibfnamefont{S.}~\bibnamefont{Schanne}},\ and\ \bibinfo {author}
  {\bibfnamefont{C.}~\bibnamefont{Winkler}},\ }%
  \bibfield{journal}{%
  \bibinfo {journal} {Nature}\ }%
  \textbf{\bibinfo {volume} {451}},\ \bibinfo {pages} {159} (\bibinfo {month}
  {Jan.}\ \bibinfo {year} {2008})%
  \bibAnnoteFile{NoStop}{2008Natur.451..159W}%
\bibitem{Jean:2003ci}%
  \BibitemOpen
  \bibfield{author}{%
  \bibinfo {author} {\bibfnamefont{P.}~\bibnamefont{Jean}} \emph{et~al.},\ }%
  \bibfield{journal}{%
  \bibinfo {journal} {Astron. Astrophys.}\ }%
  \textbf{\bibinfo {volume} {407}},\ \bibinfo {pages} {L55} (\bibinfo {year}
  {2003}),\
  \Eprint{http://arxiv.org/abs/arXiv:astro-ph/0309484}{arXiv:astro-ph/0309484}%
  \bibAnnoteFile{NoStop}{Jean:2003ci}%
\bibitem{Strong:2004de}%
  \BibitemOpen
  \bibfield{author}{%
  \bibinfo {author} {\bibfnamefont{A.~W.}\ \bibnamefont{Strong}}, \bibinfo
  {author} {\bibfnamefont{I.~V.}\ \bibnamefont{Moskalenko}},\ and\ \bibinfo
  {author} {\bibfnamefont{O.}~\bibnamefont{Reimer}},\ }%
  \bibfield{journal}{%
  \bibinfo {journal} {Astrophys. J.}\ }%
  \textbf{\bibinfo {volume} {613}},\ \bibinfo {pages} {962} (\bibinfo {year}
  {2004}),\
  \Eprint{http://arxiv.org/abs/arXiv:astro-ph/0406254}{arXiv:astro-ph/0406254}%
  \bibAnnoteFile{NoStop}{Strong:2004de}%
\bibitem{Weidenspointner:2008zl}%
  \BibitemOpen
  \bibfield{author}{%
  \bibinfo {author} {\bibfnamefont{G.}~\bibnamefont{Weidenspointner}}, \bibinfo
  {author} {\bibfnamefont{G.}~\bibnamefont{Skinner}}, \bibinfo {author}
  {\bibfnamefont{P.}~\bibnamefont{Jean}}, \bibinfo {author}
  {\bibfnamefont{J.}~\bibnamefont{Kn{\"o}dlseder}}, \bibinfo {author}
  {\bibfnamefont{P.}~\bibnamefont{von Ballmoos}}, \bibinfo {author}
  {\bibfnamefont{R.}~\bibnamefont{Diehl}}, \bibinfo {author}
  {\bibfnamefont{A.}~\bibnamefont{Strong}}, \bibinfo {author}
  {\bibfnamefont{B.}~\bibnamefont{Cordier}}, \bibinfo {author}
  {\bibfnamefont{S.}~\bibnamefont{Schanne}},\ and\ \bibinfo {author}
  {\bibfnamefont{C.}~\bibnamefont{Winkler}},\ }%
  \bibfield{journal}{%
  \bibinfo {journal} {New Astronomy Reviews}\ }%
  \textbf{\bibinfo {volume} {52}},\ \bibinfo {pages} {454 } (\bibinfo {year}
  {2008}),\ ISSN \bibinfo {issn} {1387-6473},\ \bibinfo {note} {astronomy with
  Radioactivities. VI - Proceedings of International Workshop Held at Ringberg
  Castle of Max Planck Gesellschaft in Kreuth, Germany, 7-10 January 2008}%
  \bibAnnoteFile{NoStop}{Weidenspointner:2008zl}%
\bibitem{Porter:2008fk}%
  \BibitemOpen
  \bibfield{author}{%
  \bibinfo {author} {\bibfnamefont{T.~A.}\ \bibnamefont{Porter}}, \bibinfo
  {author} {\bibfnamefont{I.~V.}\ \bibnamefont{Moskalenko}}, \bibinfo {author}
  {\bibfnamefont{A.~W.}\ \bibnamefont{Strong}}, \bibinfo {author}
  {\bibfnamefont{E.}~\bibnamefont{Orlando}},\ and\ \bibinfo {author}
  {\bibfnamefont{L.}~\bibnamefont{Bouchet}},\ }%
  \bibfield{journal}{%
  \bibinfo {journal} {Astrophys. J.}\ }%
  \textbf{\bibinfo {volume} {682}},\ \bibinfo {pages} {400} (\bibinfo {year}
  {2008}),\ \Eprint{http://arxiv.org/abs/arXiv:0804.1774}{arXiv:0804.1774}%
  \bibAnnoteFile{NoStop}{Porter:2008fk}%
\bibitem{Jaikumar:2004rp}%
  \BibitemOpen
  \bibfield{author}{%
  \bibinfo {author} {\bibfnamefont{P.}~\bibnamefont{Jaikumar}}, \bibinfo
  {author} {\bibfnamefont{C.}~\bibnamefont{Gale}}, \bibinfo {author}
  {\bibfnamefont{D.}~\bibnamefont{Page}},\ and\ \bibinfo {author}
  {\bibfnamefont{M.}~\bibnamefont{Prakash}},\ }%
  \bibfield{journal}{%
  \bibinfo {journal} {Phys. Rev. D}\ }%
  \textbf{\bibinfo {volume} {70}},\ \bibinfo {pages} {023004} (\bibinfo {month}
  {July}\ \bibinfo {year} {2004}),\
  \Eprint{http://arxiv.org/abs/arXiv:astro-ph/0403427}{arXiv:astro-ph/0403427}%
  \bibAnnoteFile{NoStop}{Jaikumar:2004rp}%
\bibitem{Caron:2009zi}%
  \BibitemOpen
  \bibfield{author}{%
  \bibinfo {author} {\bibfnamefont{J.-F.}\ \bibnamefont{Caron}}\ and\ \bibinfo
  {author} {\bibfnamefont{A.~R.}\ \bibnamefont{Zhitnitsky}},\ }%
  \bibfield{journal}{%
  \Doi{10.1103/PhysRevD.80.123006}{\bibinfo {journal} {Phys. Rev. D}}\ }%
  \textbf{\bibinfo {volume} {80}},\ \bibinfo {pages} {123006} (\bibinfo {month}
  {Dec.}\ \bibinfo {year} {2009}),\
  \Eprint{http://arxiv.org/abs/arXiv:0907.4715 [astro-ph.HE]}{arXiv:0907.4715
  [astro-ph.HE] [astro-ph.HE]}%
  \bibAnnoteFile{NoStop}{Caron:2009zi}%
\bibitem{Charbonneau:2009ax}%
  \BibitemOpen
  \bibfield{author}{%
  \bibinfo {author} {\bibfnamefont{J.}~\bibnamefont{Charbonneau}}\ and\
  \bibinfo {author} {\bibfnamefont{A.~R.}\ \bibnamefont{Zhitnitsky}},\ }%
  \bibfield{journal}{%
  \Doi{10.1088/1475-7516/2010/08/010}{\bibinfo {journal} {\textsc{jcap}}}\ }%
  \textbf{\bibinfo {volume} {2010}},\ \bibinfo {pages} {010} (\bibinfo {month}
  {Aug.}\ \bibinfo {year} {2010}),\
  \Eprint{http://arxiv.org/abs/arXiv:0903.4450}{arXiv:0903.4450 [astro-ph.HE]}%
  \bibAnnoteFile{NoStop}{Charbonneau:2009ax}%
\bibitem{Urban:2009vy}%
  \BibitemOpen
  \bibfield{author}{%
  \bibinfo {author} {\bibfnamefont{F.~R.}\ \bibnamefont{Urban}}\ and\ \bibinfo
  {author} {\bibfnamefont{A.~R.}\ \bibnamefont{Zhitnitsky}},\ }%
  \bibfield{journal}{%
  \Doi{10.1016/j.physletb.2010.03.080}{\bibinfo {journal} {Phys. Lett. B}}\ }%
  \textbf{\bibinfo {volume} {688}},\ \bibinfo {pages} {9 } (\bibinfo {month}
  {Apr.}\ \bibinfo {year} {2010}),\
  \Eprint{http://arxiv.org/abs/arXiv:0906.2162}{arXiv:0906.2162 [gr-qc]}%
  \bibAnnoteFile{NoStop}{Urban:2009vy}%
\bibitem{Urban:2009yg}%
  \BibitemOpen
  \bibfield{author}{%
  \bibinfo {author} {\bibfnamefont{F.~R.}\ \bibnamefont{Urban}}\ and\ \bibinfo
  {author} {\bibfnamefont{A.~R.}\ \bibnamefont{Zhitnitsky}},\ }%
  \bibfield{journal}{%
  \Doi{10.1016/j.nuclphysb.2010.04.001}{\bibinfo {journal} {Nucl. Phys. B}}\ }%
  \textbf{\bibinfo {volume} {835}},\ \bibinfo {pages} {135 } (\bibinfo {month}
  {Apr.}\ \bibinfo {year} {2010}),\
  \Eprint{http://arxiv.org/abs/arXiv:0909.2684}{arXiv:0909.2684 [astro-ph.CO]}%
  \bibAnnoteFile{NoStop}{Urban:2009yg}%
\bibitem{HK:1964}%
  \BibitemOpen
  \bibfield{author}{%
  \bibinfo {author} {\bibfnamefont{P.}~\bibnamefont{Hohenberg}}\ and\ \bibinfo
  {author} {\bibfnamefont{W.}~\bibnamefont{Kohn}},\ }%
  \bibfield{journal}{%
  \bibinfo {journal} {Phys. Rev.}\ }%
  \textbf{\bibinfo {volume} {136}},\ \bibinfo {pages} {B864} (\bibinfo {month}
  {Nov}\ \bibinfo {year} {1964})%
  \bibAnnoteFile{NoStop}{HK:1964}%
\bibitem{Braaten:1992rz}%
  \BibitemOpen
  \bibfield{author}{%
  \bibinfo {author} {\bibfnamefont{E.}~\bibnamefont{Braaten}},\ }%
  \bibfield{journal}{%
  \bibinfo {journal} {ApJ}\ }%
  \textbf{\bibinfo {volume} {392}},\ \bibinfo {pages} {70} (\bibinfo {month}
  {Jun.}\ \bibinfo {year} {1992})%
  \bibAnnoteFile{NoStop}{Braaten:1992rz}%
\bibitem{LS:1973}%
  \BibitemOpen
  \bibfield{author}{%
  \bibinfo {author} {\bibfnamefont{E.~H.}\ \bibnamefont{Lieb}}\ and\ \bibinfo
  {author} {\bibfnamefont{B.}~\bibnamefont{Simon}},\ }%
  \bibfield{journal}{%
  \bibinfo {journal} {Phys. Rev. Lett.}\ }%
  \textbf{\bibinfo {volume} {31}},\ \bibinfo {pages} {681} (\bibinfo {month}
  {Sep}\ \bibinfo {year} {1973})%
  \bibAnnoteFile{NoStop}{LS:1973}%
\bibitem{Prantzos:2010}%
  \BibitemOpen
  \bibfield{author}{%
  \bibinfo {author} {\bibfnamefont{N.}~\bibnamefont{Prantzos}}, \bibinfo
  {author} {\bibfnamefont{C.}~\bibnamefont{Boehm}}, \bibinfo {author}
  {\bibfnamefont{A.~M.}\ \bibnamefont{Bykov}}, \bibinfo {author}
  {\bibfnamefont{R.}~\bibnamefont{Diehl}}, \bibinfo {author}
  {\bibfnamefont{K.}~\bibnamefont{Ferriere}}, \bibinfo {author}
  {\bibfnamefont{N.}~\bibnamefont{Guessoum}}, \bibinfo {author}
  {\bibfnamefont{P.}~\bibnamefont{Jean}}, \bibinfo {author}
  {\bibfnamefont{J.}~\bibnamefont{Knoedlseder}}, \bibinfo {author}
  {\bibfnamefont{A.}~\bibnamefont{Marcowith}}, \bibinfo {author}
  {\bibfnamefont{I.~V.}\ \bibnamefont{Moskalenko}}, \bibinfo {author}
  {\bibfnamefont{A.}~\bibnamefont{Strong}},\ and\ \bibinfo {author}
  {\bibfnamefont{G.}~\bibnamefont{Weidenspointner}}}%
   (\bibinfo {year} {2010}),\
  \Eprint{http://arxiv.org/abs/arXiv:1009.4620}{arXiv:1009.4620}%
  \bibAnnoteFile{NoStop}{Prantzos:2010}%
\end{thebibliography}
\end{document}